\documentclass[12pt,journal,draftclsnofoot,oneside,onecolumn]{IEEEtran}

\usepackage[cmex10]{amsmath}
\usepackage{amssymb}
\usepackage{epsfig}
\usepackage{subfigure}
\usepackage{cite}
\usepackage{graphicx}
\usepackage{color}
\usepackage{bm}
\usepackage{booktabs}
\usepackage[cmex10]{amsmath}

\newtheorem{lemma}{\it Lemma}

\newtheorem{remark}{\it Remark}
\newtheorem{proposition}{\it Proposition}
\begin{document}

\title{Securing NOMA Networks by Exploiting Intelligent Reflecting Surface}
\author{Zheng Zhang, Jian Chen,~\IEEEmembership{Member,~IEEE}, Qingqing Wu,~\IEEEmembership{Member,~IEEE}, \\ Yuanwei Liu,~\IEEEmembership{Senior Member,~IEEE}, Lu Lv,~\IEEEmembership{Member,~IEEE}, and Xunqi Su, 
\thanks{Zheng Zhang, Jian Chen, Lu Lv, and Xunqi Su are with the State Key Laboratory of Integrated Services Networks, Xidian University, Xi'an 710071, China (e-mail: zzhang\_688@stu.xidian.edu.cn; jianchen@mail.xidian.edu.cn; lulv@xidian.edu.cn; xunqisu@stu.xidian.edu.cn;). Qingqing Wu is with the State Key Laboratory of Internet of Things for Smart City, University of Macau, Macau 999078, China (e-mail: qingqingwu@um.edu.mo). Yuanwei Liu is with the School of Electronic Engineering and Computer Science, Queen Mary University of London, London E1 4NS, U.K. (e-mail: yuanwei.liu@qmul.ac.uk).}
}
\maketitle
\begin{abstract}

    This paper investigates the security enhancement of an intelligent reflecting surface (IRS) assisted non-orthogonal multiple access (NOMA) network, where a distributed IRS enabled NOMA transmission framework is proposed to serve users securely in the presence of a passive eavesdropper. Considering that eavesdropper's instantaneous channel state information (CSI) is challenging to acquire in practice, we utilize secrecy outage probability (SOP) as the security metric. A problem of maximizing the minimum secrecy rate among users, subject to the successive interference cancellation (SIC) decoding constraints and SOP constraints, by jointly optimizing transmit beamforming at the BS and phase shifts of IRSs, is formulated. For special case with a single-antenna BS, we derive the exact closed-form SOP expressions and propose a novel \textit{ring-penalty} based successive convex approximation (SCA) algorithm to design power allocation and phase shifts jointly. While for the more general and challenging case with a multi-antenna BS, we adopt the Bernstein-type inequality to approximate the SOP constraints by a deterministic convex form. To proceed, an efficient alternating optimization (AO) algorithm is developed to solve the considered problem. Numerical results validate the advantages of the proposed algorithms over the baseline schemes. Particularly, two interesting phenomena on distributed IRS deployment are revealed: 1) the secrecy rate peak is achieved only when distributed IRSs share the reflecting elements equally; and 2) the distributed IRS deployment does not always outperform the centralized IRS deployment, due to the tradeoff between the number of IRSs and the reflecting elements equipped at each IRS.

\end{abstract}

\IEEEpeerreviewmaketitle

\section{Introduction}\label{Introduction}
Over the past decade, a variety of key technologies have been spawned for fifth-generation (5G) wireless communication, such as massive multiple-input multiple-output (MIMO) network, ultra-dense network (UDN) and millimeter wave (mmWave) network \cite{F.Boccardi_5G_magazine}. Although these 5G-oriented key enablers have demonstrated their tremendous potential in achieving massive connectivity, high spectrum efficiency and low latency, there inherent limitations of high energy consumption and hardware complexity are still critical challenges in practice, which thus motivates both the academia and industry to find a green and cost-effective solution for future wireless networks. Recently, a novel energy-efficient technique, namely, intelligent reflecting surface (IRS), has received significant attention due to its ability to control electromagnetic waves \cite{R.Zhang_IRS_magazine}. Generally, IRS is a kind of metasurface consisting of a large number of passive tunable reflecting elements \cite{W.Tang_RIS}, each of which can independently adjust the amplitudes and phase shifts of the reflected signals, thereby achieving an active control of the radio propagation environment. Compared with the existing active relay equipped with massive antennas, IRS is much more energy-efficient and less costly, as it alters the reflection of signals without requiring any active module and is capable of providing an appealing \textit{squared} array power gain with the growing number of reflecting elements \cite{R.Zhang_IRS_passive,R.Zhang_IRS_discrete}.

Furthermore, IRS also shows good adaptability, which can be integrated into assorted scenarios. In particular, as IRS can artificially create differences between the users' effective/combined channels, the integration of IRS into non-orthogonal multiple access (NOMA) networks can provide an appealing performance improvement in energy efficiency, spectrum utilization and user fairness \cite{Y.Liu_NOMA_RIS}, which thus has received increasing  attention \cite{B.Zheng_OMA_NOMA,X.Mu_IRS_NOMA,J.Zuo_IRS_NOMA,Z.Ding_IRS_NOMA_1,Z.Ding_IRS_NOMA_2,J.Zhu_IRS_NOMA}.
Specifically, the transmit power consumption performances of IRS assisted NOMA and orthogonal multiple access (OMA) networks were analyzed and compared in \cite{B.Zheng_OMA_NOMA}. To further improve the spectrum efficiency, the joint active/passive beamforming optimization problem was investigated in \cite{X.Mu_IRS_NOMA} and \cite{J.Zuo_IRS_NOMA}. On the other hand, for the orthogonal channel scenario, a spatial division multiple access (SDMA) based IRS assisted NOMA scheme was proposed in \cite{Z.Ding_IRS_NOMA_1}. The impact of coherent phase shifting and random discrete phase shifting on the IRS assisted NOMA communication was studied in \cite{Z.Ding_IRS_NOMA_2}, which revealed their tradeoff between reliability and complexity. Additionally, a novel transmission scheme for multiple-input single-output (MISO) IRS assisted NOMA communications was designed in \cite{J.Zhu_IRS_NOMA} from the energy-efficient viewpoint.

However, due to the broadcast characteristic of wireless channels, any user (even a malicious eavesdropper) is capable of accessing the wireless network with no difficulty, which exposes the private data to a vulnerable communication environment. To address this, physical layer (information-theoretic) security (PLS) is developed to secure legitimate communications via exploiting the characteristics of wireless channels, such as interference, fading, and noise, which efficiently avoids the complex encryption algorithm design and secret key management in the upper layers \cite{X.Zhou_PLS}. Take NOMA networks as an example, many works have utilized the PLS to secure multi-user communications \cite{Y.Zhang_NOMA_PLS,B.He_SOP,L.Lv_PLS_NOMA_Untrusted,K.Cao_NOMA_PLS,H.-M.Wang_NOMA_PLS,L.Lv_NOMA_magazine2}. In \cite{Y.Zhang_NOMA_PLS,B.He_SOP}, the secrecy capacity maximization problems of single-input single-output (SISO) NOMA networks were investigated, in which the optimal power allocation policies for different objectives (i.e., sum-secrecy rate and fairness-secrecy rate) were derived. In the untrusted relay scenario, a cooperative secrecy scheme of both uplink and downlink NOMA cases was proposed in \cite{L.Lv_PLS_NOMA_Untrusted}. While in the untrusted user scenario, two optimal relay selection schemes were developed in \cite{K.Cao_NOMA_PLS}. The authors of \cite{H.-M.Wang_NOMA_PLS} studied the joint beamforming optimization problem with the aid of artificial noise (AN) for jamming eavesdroppers. Furthermore, a new interference exploitation based jamming strategy is proposed to enhance security of NOMA networks in \cite{L.Lv_NOMA_magazine2}. Thanks to the IRS's ability of reconfiguring wireless channels intelligently, it is expected to further improve the PLS performance \cite{M.Cui_IRS_PLS,Z.Chu_IRS_Security,X.Guan_IRS_AN,X.Yu_imperfectCSI,L.Dong_Without_CSI,S.Hong_IRS_MIMO,L.Lv_IRS_Two-Way,ZZ_Robust,N.Li_IRS_NOMA_Security}.
The authors of \cite{M.Cui_IRS_PLS} and \cite{Z.Chu_IRS_Security} first studied the possibility of security improvement by integrating IRS into wireless networks, which confirmed the huge potential of IRS assisted PLS communication. In \cite{X.Guan_IRS_AN}, the author explored the influence of the AN on the IRS assisted secure communication. Considering the passivity of eavesdroppers, a robust security-enhancing scheme against imperfect channel state information (CSI) eavesdroppers was proposed in \cite{X.Yu_imperfectCSI}. The authors of \cite{L.Dong_Without_CSI} further investigated the scenario without eavesdropper's CSI. While for MIMO networks, the IRS assisted secure wireless transmission was studied in \cite{S.Hong_IRS_MIMO}. A novel IRS assisted jamming scheme was proposed in \cite{L.Lv_IRS_Two-Way} for two-way communication secrecy enhancement. More recently, the authors of \cite{ZZ_Robust} and \cite{N.Li_IRS_NOMA_Security} have studied the secure transmission problem of IRS assisted NOMA networks, which demonstrates the great potential of IRS in security enhancement of NOMA communication.

\subsection{Motivations and Contributions}
From the aforementioned works \cite{M.Cui_IRS_PLS,Z.Chu_IRS_Security,X.Guan_IRS_AN,X.Yu_imperfectCSI,L.Dong_Without_CSI,S.Hong_IRS_MIMO,L.Lv_IRS_Two-Way}, it is known that by designing the reflection amplitude/phase shift appropriately, IRS is capable of bringing significant security enhancement to wireless networks. However, their results are not applicable to the case with NOMA since the successive interference cancellation (SIC) decoding was not taken into account. In NOMA networks, IRS also needs to achieve the tradeoff between guaranteeing the successful SIC decoding and the user channel strengthening/suppressing, because the SIC decoding usually limits the channel strength of the user with higher decoding priority, which may result in that some legitimate channels are weaker than eavesdropping channels and lead to degraded network security. Although there were a handful of works devoted to the secure IRS assisted NOMA transmissions \cite{ZZ_Robust,N.Li_IRS_NOMA_Security}, only the instantaneous eavesdropping CSI available scenario was considered. Unfortunately, acquiring the instantaneous CSI of a passive eavesdropper is much difficult in practice since it tends to hide itself from legitimate nodes. If the IRS just possesses the knowledge of eavesdroppers' statistical CSI rather than instantaneous CSI, it naturally weakens its ability to degrade the eavesdropping channels. Instead, the IRS can only enhance the signal reception of NOMA users according to their instantaneous CSI, which, however, may also benefit the wiretapping. Therefore, a fundamental issue appears: \textit{how to utilize IRS to secure NOMA transmissions against the passive eavesdropper with only its statistical CSI?} To our best knowledge, this question has not been addressed in the literature.

Motivated by the above, we focus on the secure NOMA transmission without instantaneous CSI of the eavesdropper, where the joint optimization schemes regarding to  the transmit power/beamforming at the base station (BS) and the reflection coefficients of IRS are developed to enhance the secrecy performance of NOMA users. Specifically, our main contributions are summarized as follows.
\begin{itemize}
  \item We propose an IRS assisted NOMA transmission framework against the passive eavesdropper, where distributed IRSs are deployed near users to prevent information leakage and improve the legitimate reception quality. Considering that only the statistical CSI of the Eve is available, we utilize the secrecy outage probability (SOP) as the security metric. Accordingly, we formulate a joint transmit beamforming and reflection coefficients design problem to maximize the minimum secrecy rate among legitimate users, subject to the total transmit power constraint at the BS, the phase shifts constraints of IRSs, the SIC decoding constraints, and the SOP constraints.
  \item To handle the non-convex and challenging optimization problem, we first consider the special case with a single-antenna BS. In this case, we derive the exact SOP of each user in closed-form expression, and the result indicates that reception quality of eavesdropper is only related to power allocation at the BS but is independent of phase shifts of IRSs. To enhance the signal strength and prevent the information leakage at legitimate users, we develop a ring-penalty based successive convex approximation (SCA) algorithm to optimize transmit power allocation and phase setting jointly, where the SCA technique is used to decouple the optimization variables while the ring-penalty method is proposed to relax the rank-one constraint. 
  \item Next, we investigate the general case with a multi-antenna BS. Since the SOP constraints have no closed-form expressions, we define the joint beamforming matrix and apply the Bernstein-type inequality to obtain a conservative approximation form, which implies that even without eavesdroppers' instantaneous CSI, IRSs can still suppress the eavesdropping channel condition efficiently. Then, an alternating optimization (AO) algorithm is proposed to optimize the transmit beamforming at the BS and reflection coefficients of IRSs alternately, where a difference-of-convex relaxation (DCR) based Dinkelbach algorithm is designed to obtain the rank-one transmit beamforming matrix optimally, while the modified ring-penalty based SCA algorithm is developed to search the optimal phase shifts of IRSs.
  \item Numerical results validate the advantages of the proposed scheme in comparison to other baseline schemes. Particularly, we draw two interesting insights for the IRS deployment: 1) under the fixed number of the IRSs, the best secrecy performance is only achieved when distributed IRSs share the reflecting elements equally; and 2) given the total number of the reflecting elements, increasing the number of the IRSs does not necessarily lead to higher secrecy performance, and there exists a tradeoff between the number of IRSs and that of the reflecting elements equipped at each IRS.
\end{itemize}

The organization of this paper is as follows. Section \ref{fir:Model} introduces the system model and the problem formulation. In Section \ref{single_antenna_case}, we develop a ring-penalty based SCA algorithm to tackle the problem for the single-antenna BS case. Section \ref{multi_antenna_case} proposes an AO algorithm to optimize transmit beamforming and reflection coefficients jointly. The numerical results and discussions are shown in Section \ref{Simulation}. Finally, our conclusions are presented in Section \ref{conclusion}.

\textit{Notations:} boldface capital $\mathbf{Z}$ and lower-case letter $\mathbf{z}$ denote matrix and vector respectively. For the complex-valued matrix $\mathbf{Z}$, $\mathbf{Z}^{T}$ and $\mathbf{Z}^{H}$ denote transpose and Hermitian conjugate operations, while $\text{rank}(\mathbf{Z})$, $\text{Tr}(\mathbf{Z})$ and $\|\mathbf{Z}\|_{2}$ denote rank value, trace operation and spectral norm. $\text{diag}(\mathbf{z})$ and $\text{blkdiag}(\mathbf{[\mathbf{Z}_{1},\dots,\mathbf{Z}_{n}]})$ represent diagonal and block diagonal operations. $\mathbb{E}($\textperiodcentered$)$ and $\mathbb{P}($\textperiodcentered$)$ are the statistical expectation and probability, respectively. $\mathbf{I}$ is the identity matrix. $\Re(\cdot)$ denotes the real component of the complex value. $\rho_{\text{max}}(\mathbf{Z})$ denotes extracting maximal eigenvalue operation, while $\rho_{i}$ and $\sigma_{i}$ respectively present the \textit{i}th largest eigenvalue and singular value of corresponding matrix unless otherwise specified. $\mathbf{Z}\succeq\mathbf{0}$ means that $\mathbf{Z}$ is a positive semidefinite matrix, while $\mathbf{z}\sim \mathcal{CN}(0,\mathbf{Z})$ denotes a circularly symmetric complex Gaussian (CSCG) vector with zero mean and covariance matrix $\mathbf{Z}$.

\section{System Model and Problem Formulation}\label{fir:Model}

\subsection{System Setup}\label{fir:system model}
\begin{figure}
  \centering
  \includegraphics[scale = 0.4]{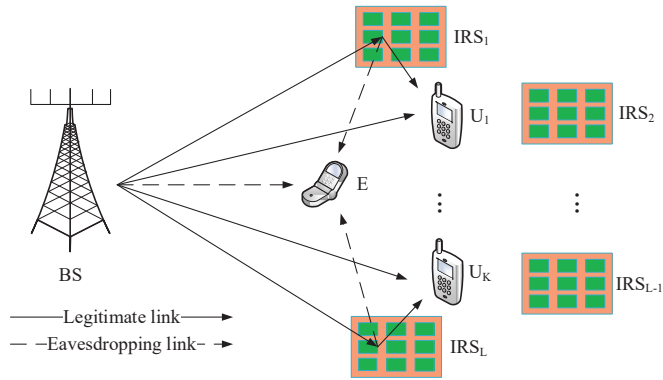}
  \caption{A distributed IRS assisted secure NOMA network.}
  \label{Fig.1}
\end{figure}

We consider the secure downlink communication of an IRS-assisted NOMA network as shown in Fig. \ref{Fig.1}, which consists of a BS, $K$ ($K\geq1$) legitimate users (denoted by $\text{U}_{i}$, $i\in\{1,\dots,K\}$), one eavesdropper (E) and $L$ ($L\geq1$) IRSs (denoted by $\text{IRS}_{l}$, $l\in\{1,\dots,L\}$). To protect the superposed signals intended for NOMA users against malicious eavesdropping, distributed IRSs are deployed near the receivers to reduce information leakage and strengthen the signal power at legitimate users simultaneously. We assume that the BS is equipped with $M$ ($M\geq1$) antennas, while the legitimate users and eavesdropper are equipped with single antenna. It is also assumed that a total number of $N$ reflecting elements are shared by $L$ IRSs, and thus, we have $\sum_{l=1}^{L}{N_{l}}=N$, where $N_{l}$ denotes the number of reflecting elements of $\text{IRS}_{l}$. In the considered network, only the first-order reflected signals are taken into consideration since the multiple reflection signals suffer from severe path loss, which can be neglected reasonably \cite{R.Zhang_IRS_magazine,R.Zhang_IRS_passive}. Furthermore, a smart controller is attached to each of the IRS, which communicates with the BS via a separate wireless link for coordinating transmission and exchanging information, e.g. channel knowledge, and controls the phase shifts of all reflecting elements in real time.

All the channels are assumed to be quasi-static block fading channels, which means that the channel coefficients remain constant in one fading block but can change independently across different fading blocks. The baseband equivalent channels from the BS to $\text{IRS}_{l}$, $\text{U}_{i}$ and E are denoted by $\mathbf{H}_{\text{B},\text{I}_{l}}\in\mathbb{C}^{N_{l}\times M}$, $\mathbf{h}_{\text{B},i}\in\mathbb{C}^{M\times 1}$ and $\mathbf{h}_{\text{B},\text{E}}\in\mathbb{C}^{M\times 1}$, while the channel coefficients from $\text{IRS}_{l}$ to $\text{U}_{i}$ and E are represented by $\mathbf{h}_{\text{I}_{l},i}\in\mathbb{C}^{N_{l}\times 1}$ and $\mathbf{h}_{\text{I}_{l},\text{E}}\in\mathbb{C}^{N_{l}\times 1}$. Note that there exist two different links in the considered network. Specifically, since the IRSs and BS usually have fixed positions and can be properly deployed to favor line-of-sight (LoS) transmissions in practice, we assume the Rician fading model for the BS-$\text{IRS}_{l}$ links, i.e.,
\begin{equation}
\label{1}
\mathbf{H}_{\text{B},\text{I}_{l}}=L(d)\left(\sqrt{\frac{\kappa}{1+\kappa}}\mathbf{\hat{H}}_{\text{B},\text{I}_{l}}+\sqrt{\frac{1}
{1+\kappa}}\mathbf{\tilde{H}}_{\text{B},\text{I}_{l}}\right),
\end{equation}
where $L(d)$ denotes large-scale path loss, $\kappa$ denotes the Rician factor, and $\mathbf{\hat{H}}_{\text{B},\text{I}_{l}}$ and $\mathbf{\tilde{H}}_{\text{B},\text{I}_{l}}$ denote the LoS and non-line-of-sight (NLoS) components respectively. $\mathbf{\hat{H}}_{\text{B},\text{I}_{l}}$ is modeled as the product of the steering vectors of the antenna array of transceivers, while $\mathbf{\tilde{H}}_{\text{B},\text{I}_{l}}$ is Rayleigh fading \cite{S.Hong_IRS_MIMO,X.Yu_imperfectCSI}. On the other hand, considering the mobility of the receiving nodes, the LoS links from the BS/IRSs to the receivers may not exist. Hence, we assume the Rayleigh fading for the remaining links, i.e., $\mathbf{h}\sim \mathcal{CN}(0,L(d)^{2}\mathbf{I})$, where $\mathbf{h}\in\{\mathbf{h}_{\text{B},i},\mathbf{h}_{\text{B},\text{E}}, \mathbf{h}_{\text{I}_{l},i},\mathbf{h}_{\text{I}_{l},\text{E}}\}$. The
large-scale fading can be expressed as $L(d)=\sqrt{L_{0}d^{-{\alpha}}}$, where $L_{0}$ represents the path loss at the reference distance of $1$ meter, $d$ denotes the distance between transceivers, and $\alpha$ denotes the corresponding path-loss exponent. Additionally, we denote the reflection coefficients of $n$th element as $\beta_{n}e^{j\alpha_{n}}$, where $\alpha_{n}\in[0,2\pi)$ and $\beta_{n}\in[0,1]$. For the ease of practical hardware implementation, we assume maximum reflection amplitude for each element, i.e., $\beta_{n}=1$ \cite{R.Zhang_IRS_magazine,R.Zhang_IRS_passive}. As a result, the reflection coefficients matrix of $\text{IRS}_{l}$ can be given by $\bm{\Theta}_{l} = \text{diag}([e^{j\alpha_{1}},\dots,
e^{j\alpha_{N_{l}}}]^{T})\in\mathbb{C}^{N_{l}\times N_{l}}$.

In this paper, we assume that the instantaneous CSI of the legitimate channels are perfectly known at the BS, which can be achieved by the simultaneous-user channel estimation (SiUCE) scheme \cite{B.Zheng_channel_estimation_1} or the pilot-based channel estimation method \cite{Guan_channel_estimation,Z.Wang_channel_estimation}. However, the acquisition of the instantaneous CSI for eavesdropper is difficult to obtain in practice since the eavesdropper tends to keep silent, and does not exchange any information with the BS when wiretapping the legitimate communications. Therefore, we assume that the BS only possesses the channel statistics of E, which can be estimated by the fading knowledge and average distance between transceivers \cite{B.He_SOP}. On the other hand, considering the fact that the eavesdropper can intercept signals from the BS to estimate CSI between BS and itself, and thus, we assume that E knows its own instantaneous CSI perfectly, which is also the worse-case setup and serves as the benchmark scheme for other assumptions.

\subsection{Transmission Scheme}
To serve multiple users with the same time-frequency resource block, the BS transmits superimposed signals by exploiting multiple beamforming vectors, i.e., $\mathbf{s}=\sum_{i=1}^{K}\mathbf{w}_{i}s_{i}$, where $s_{i}$ denotes the target signal of $\text{U}_{i}$ with $\mathbb{E}\{|s_{i}|^{2}\}=1$, and $\mathbf{w}_{i}\in\mathbb{C}^{M\times 1}$ denotes the corresponding vector. Accordingly, the received signals at $\text{U}_{i}$ and $\text{E}$ are given, respectively, by
\begin{align}
\label{2}\nonumber
y_{i}=&\left(\sum\nolimits_{l=1}^{L}\mathbf{h}_{\text{I}_{l},i}^{H}\bm{\Theta}_{l}\mathbf{H}_{\text{B},\text{I}_{l}}
+\mathbf{h}_{\text{B},i}^{H}\right)\sum\nolimits_{i=1}^{K}\mathbf{w}_{i}s_{i}+n_{i}\\
=&\left(\mathbf{h}_{\text{I},i}^{H}\bm{\Theta}\mathbf{H}_{\text{B},\text{I}}
+\mathbf{h}_{\text{B},i}^{H}\right)\sum\nolimits_{i=1}^{K}\mathbf{w}_{i}s_{i}+n_{i},
\end{align}
\begin{align}
\label{3}\nonumber
y_{\text{E}}=&\left(\sum\nolimits_{l=1}^{L}\mathbf{h}_{\text{I}_{l},\text{E}}^{H}\bm{\Theta}_{l}\mathbf{H}_{\text{B},\text{I}_{l}}
+\mathbf{h}_{\text{B},\text{E}}^{H}\right)\sum\nolimits_{i=1}^{K}\mathbf{w}_{i}s_{i}+n_{\text{E}}\\
=&\left(\mathbf{h}_{\text{I},\text{E}}^{H}\bm{\Theta}\mathbf{H}_{\text{B},\text{I}}
+\mathbf{h}_{\text{B},\text{E}}^{H}\right)\sum\nolimits_{i=1}^{K}\mathbf{w}_{i}s_{i}+n_{\text{E}},
\end{align}
where $n_{i}$ and $n_{\text{E}}$ represent the additive white Gaussian noise (AWGN) at $\text{U}_{i}$ and $\text{E}$ with zero mean and variance $\sigma^{2}$, respectively, while $\mathbf{h}_{\text{I},i}^{H}=[\mathbf{h}_{\text{I}_{1},i}^{H},\dots,\mathbf{h}_{\text{I}_{L},i}^{H}]$, $\mathbf{h}_{\text{I},\text{E}}^{H}=[\mathbf{h}_{\text{I}_{1},\text{E}}^{H},\dots,
\mathbf{h}_{\text{I}_{L},\text{E}}^{H}]$, $\bm{\Theta}=\text{blkdiag}(\bm{\Theta}_{1},\dots,\bm{\Theta}_{L})$ and  $\mathbf{h}_{\text{B},\text{I}} = [\mathbf{h}_{\text{B},\text{I}_{1}},\dots,\mathbf{h}_{\text{B},\text{I}_{L}}]^{T}$. 


In IRS assisted NOMA networks, each receiver adopts SIC technique to detect superimposed signals according to the equivalent reconfigurable channel (include direct and cascade channels) qualities and beamforming vectors \cite{Y.Liu_NOMA_RIS}. Define the decoding order map $\lambda(j)=i$, with indicating that the signals of $\text{U}_{i}$ are decoded at the $j$th stage of SIC. More specifically, $\text{U}_{\lambda(j)}$ first decodes the signals of $\text{U}_{\lambda(j-m)}$ ($0<m<j\leq K$), and removes these signals from its decoding results. Then, it decodes its own signal by treating signals for $\text{U}_{\lambda(j+n)}$ ($0<n\leq K-j$) as co-channel interference. For convenience of exposition, we consider the fixed decoding order, which satisfies $\lambda(i)=i$ \cite{X.Mu_SIC}. As such, the decoding order at $\text{U}_{i}$ is given by $s_{1}\rightarrow\dots \rightarrow s_{i}$. Note that the achievable rate at $\text{U}_{k}$ to decode $s_{i}$ should be no less than the achievable rate at $\text{U}_{i}$ to decode $s_{i}$ ($1\leq i\leq k\leq K$) for guaranteeing successful SIC decoding \cite{Y.Liu_MIMO_NOMA}. Also, to balance the user fairness, more power should be allocated to the weaker channel users, i.e., $|(\mathbf{h}_{\text{I},i}^{H}\bm{\Theta}\mathbf{H}_{\text{B},\text{I}}
+\mathbf{h}_{\text{B},i}^{H})\mathbf{w}_{i}|\leq\dots\leq|(\mathbf{h}_{\text{I},i}^{H}\bm{\Theta}\mathbf{H}_{\text{B},\text{I}}
+\mathbf{h}_{\text{B},i}^{H})\mathbf{w}_{1}|$ \cite{X.Mu_IRS_NOMA}.

The achievable rate at $\text{U}_{i}$ for decoding its own message is given by
\begin{equation}
\label{4}
R_{i,i}=\log_{2}\left(1+\frac{|(\mathbf{h}_{\text{I},i}^{H}\bm{\Theta}\mathbf{H}_{\text{B},\text{I}}+\mathbf{h}_{\text{B},i}^{H})\mathbf{w}_{i}|^{2}}
{\sum_{j=i+1}^{K}|(\mathbf{h}_{\text{I},i}^{H}\bm{\Theta}\mathbf{H}_{\text{B},\text{I}}+\mathbf{h}_{\text{B},i}^{H})\mathbf{w}_{j}|^{2}+\sigma^{2}}\right).
\end{equation}
As for E, we further assume that E perfectly knows the decoding order and the precoding vector information, so that it can carry out SIC to detect the target signals similar to the legitimate users. Thus, the achievable rate of $\text{E}$ to decode $s_{i}$ is expressed as
\begin{equation}
\label{5}
R_{\text{E},i}=\log_{2}\left(1+\frac{|(\mathbf{h}_{\text{I},\text{E}}^{H}\bm{\Theta}\mathbf{H}_{\text{B},\text{I}}+\mathbf{h}_{\text{B},\text{E}}^{H})\mathbf{w}_{i}|^{2}}
{\sum_{j=i+1}^{K}|(\mathbf{h}_{\text{I},\text{E}}^{H}\bm{\Theta}\mathbf{H}_{\text{B},\text{I}}+\mathbf{h}_{\text{B},\text{E}}^{H})\mathbf{w}_{j}|^{2}+\sigma^{2}}\right).
\end{equation}

Due to lacking of instantaneous CSI of E, we consider the wiretap code \cite{B.He_SOP} and adopt the SOP as the security metric. Specifically, the positive difference between the codeword rate $R_{i,i}$ and the secrecy rate $R_{\text{s},i}$, i.e., redundant rate, is used to provide security against E, and the SOP of $\text{U}_{i}$ is defined as the probability that wiretapping capacity of E exceeds the redundant rate \cite{H.-M.Wang_NOMA_PLS}. Thus, the SOP of $\text{U}_{i}$ is given by
\begin{equation}
\label{6}
p_{\text{so},i}=\mathbb{P}(R_{\text{E},i}>R_{i,i}-R_{\text{s},i}).
\end{equation}

\subsection{Problem Formulation}
In this paper, we aim to maximize the minimum secrecy rate of legitimate users subject to the total power constraint at the BS, the phase shifts constraints of IRSs and the SOP/SIC constraints at legitimate users, by designing the BS's transmit beamforming vectors and IRSs' reflection coefficients jointly. The optimization problem is formulated as follows.

\begin{subequations}
\begin{align}
\label{7a} &\max\limits_{\bm{\Theta},\mathbf{w}_{i},R_{\text{s},{i}}}\quad \min\limits_{1\leq i\leq K}\  R_{\text{s},{i}},\\
\label{7b}&\quad\text{s.t.} \quad \sum\nolimits_{i=1}^{K}\|\mathbf{w}_{i}\|^{2}\leq P_{\text{B}},\\
\label{7c}&\quad\quad\quad\,\ |(\mathbf{h}_{\text{I},i}^{H}\bm{\Theta}\mathbf{H}_{\text{B},\text{I}}+\mathbf{h}_{\text{B},i}^{H})\mathbf{w}_{i}|\leq\dots\leq|
(\mathbf{h}_{\text{I},i}^{H}\bm{\Theta}\mathbf{H}_{\text{B},\text{I}}+\mathbf{h}_{\text{B},i}^{H})\mathbf{w}_{1}|,\quad  1\leq i\leq K,\\
\label{7d}&\quad\quad\quad\,\, R_{i,i}\leq R_{k,i}, \quad 1\leq i\leq k\leq K,\\
\label{7e}&\quad\quad\quad\,\, 0\leq \alpha_{n}\leq 2\pi,\quad 1\leq n\leq N,\\
\label{7f}&\quad\quad\quad\,\, \mathbb{P}(R_{\text{E},i}>R_{i,i}-R_{\text{s},i}) \leq p_{\text{max},i},\quad 1\leq i\leq K,
\end{align}
\end{subequations}
where $P_{\text{B}}$ denotes the maximum transmit power at the BS, and $p_{\text{max},i}$ represents the maximum tolerant SOP of $\text{U}_{i}$. In the problem (7), constraint \eqref{7b} limits the total transmit power at the BS; \eqref{7c} denotes the user fairness constraints; \eqref{7d} guarantees that SIC can be performed successfully; \eqref{7e} denotes phase shifts constraints of IRSs; and \eqref{7f} denotes the secrecy requirements of legitimate users. The optimization problem (7) is difficult to tackle due to the non-convex constraints \eqref{7c}, \eqref{7d}, \eqref{7f} and the coupled variables ($\bm{\Theta}$, $\mathbf{w}_{i}$). To efficiently solve this issue, we first investigate the special case, i.e., SISO network, in Section \ref{single_antenna_case}, where a ring-penalty based SCA algorithm is proposed. Based on it, an AO algorithm is developed in Section \ref{multi_antenna_case} to handle the general MISO case.


\section{Single-Antenna System}\label{single_antenna_case}

In this section, we consider a special system setup, i.e., single-antenna BS case, in order to obtain the optimal power/reflection coefficients solution and draw useful insights into the system design. In this case, the channel matrix $\mathbf{H}_{\text{B},\text{I}}$ reduces to the channel vector $\mathbf{h}_{\text{B},\text{I}}$, the channel vectors $\mathbf{h}_{\text{B},i}$ reduce to the channel coefficients $\text{h}_{\text{B},i}$, and the transmit beamforming $\mathbf{w}_{i}$ reduces to the transmit power $P_{i}$. Furthermore, the constraints \eqref{7c} and \eqref{7d} are equivalent to the channel order constraint, i.e., $|\text{h}_{\text{I},1}^{H}\bm{\Theta}\mathbf{h}_{\text{B},\text{I}}+\text{h}_{\text{B},1}^{H}|\leq\dots\leq|
\text{h}_{\text{I},K}^{H}\bm{\Theta}\mathbf{h}_{\text{B},\text{I}}+\text{h}_{\text{B},K}^{H}|$. As a result, the optimization problem is simplified as

\begin{subequations}
\begin{align}
\label{8a} &\max\limits_{\bm{\Theta},P_{i},R_{\text{s},{i}}}\quad \min\limits_{1\leq i\leq K}\  R_{\text{s},{i}},\\
\label{8b}&\quad\text{s.t.} \quad \sum\nolimits_{i=1}^{K}P_{i}\leq P_{\text{B}},\\
\label{8c}&\quad\quad\quad\,\ |\mathbf{h}_{\text{I},1}^{H}\bm{\Theta}\mathbf{h}_{\text{B},\text{I}}+\text{h}_{\text{B},1}^{H}|\leq\dots\leq|
\mathbf{h}_{\text{I},K}^{H}\bm{\Theta}\mathbf{h}_{\text{B},\text{I}}+\text{h}_{\text{B},K}^{H}|,\\
\label{8d}&\quad\quad\quad\,\, 0\leq \alpha_{n}\leq 2\pi,\quad1\leq n\leq N,\\
\label{8e}&\quad\quad\quad\,\, \mathbb{P}(R_{\text{E},i}>R_{i,i}-R_{\text{s},i}) \leq p_{\text{max},i},\quad 1\leq i\leq K,
\end{align}
\end{subequations}
where $R_{\text{E},i}=\log_{2}\Big(1+\frac{|\mathbf{h}_{\text{I},\text{E}}^{H}\bm{\Theta}\mathbf{h}_{\text{B},\text{I}}+\text{h}_{\text{B},\text{E}}^{H}|^{2}P_{i}}
{\sum_{j=i+1}^{K}|\mathbf{h}_{\text{I},\text{E}}^{H}\bm{\Theta}\mathbf{h}_{\text{B},\text{I}}+\text{h}_{\text{B},\text{E}}^{H}|^{2}P_{j}+\sigma^{2}}\Big)$ and $R_{i,i}=\log_{2}\Big(1+\frac{|\mathbf{h}_{\text{I},i}^{H}\bm{\Theta}\mathbf{h}_{\text{B},\text{I}}+\text{h}_{\text{B},i}^{H}|^{2}P_{i}}
{\sum_{j=i+1}^{K}|\mathbf{h}_{\text{I},i}^{H}\bm{\Theta}\mathbf{h}_{\text{B},\text{I}}+\text{h}_{\text{B},i}^{H}|^{2}P_{j}+\sigma^{2}}\Big)$. 

\subsection{Ring-penalty Based SCA Algorithm}\label{sec:REO_order}
To facilitate the expression of the combined channel, we define $\mathbf{u}=[e^{j\alpha_{1}},\dots,e^{j\alpha_{N}}]^{H}$, $\mathbf{\bar{u}}=[\mathbf{u};1]$, $\mathbf{q}_{i}=\text{diag}(\mathbf{h}_{\text{I},i}^{H})\mathbf{h}_{\text{B},\text{I}}$ and $\mathbf{q}_{\text{E}}=\text{diag}(\mathbf{h}_{\text{I},\text{E}}^{H})\mathbf{h}_{\text{B},\text{I}}$. Therefore, the quadratic term $|\mathbf{h}_{\text{I},i}^{H}\bm{\Theta}\mathbf{h}_{\text{B},\text{I}}+h_{\text{B},i}^{H}|^{2}$ and $|\mathbf{h}_{\text{I},\text{E}}^{H}\bm{\Theta}\mathbf{h}_{\text{B},\text{I}}+h_{\text{B},\text{E}}^{H}|^{2}$ can be rewritten as $\text{Tr}(\mathbf{J}_{i}\mathbf{U})+|h_{\text{B},i}^{H}|^{2}$ and $\text{Tr}(\mathbf{J}_{\text{E}}\mathbf{U})+|h_{\text{B},\text{E}}^{H}|^{2}$, in which
\begin{equation}
\label{9} \mathbf{U}=\mathbf{\bar{u}}\mathbf{\bar{u}}^{H},
\end{equation}
\begin{equation}\label{10}
 \mathbf{J}_{i}=\begin{bmatrix} \mathbf{q}_{i}\mathbf{q}_{i}^{H} & \mathbf{q}_{i}h_{\text{B},i} \\ h_{\text{B},i}^{H}\mathbf{q}_{i}^{H} & 0 \end{bmatrix}, \quad \mathbf{J}_{\text{E}}=\begin{bmatrix} \mathbf{q}_{\text{E}}\mathbf{q}_{\text{E}}^{H} & \mathbf{q}_{\text{E}}h_{\text{B},\text{E}} \\ h_{\text{B},\text{E}}^{H}\mathbf{q}_{\text{E}}^{H} & 0 \end{bmatrix}.
\end{equation}
Therefore, we can rewrite the constraints \eqref{8c} and \eqref{8d} as the convex forms, i.e.,
\begin{equation}\label{11}
\text{Tr}(\mathbf{J}_{1}\mathbf{U})+|h_{\text{B},1}^{H}|^{2}\leq\dots\leq\text{Tr}(\mathbf{J}_{K}\mathbf{U})+|h_{\text{B},K}^{H}|^{2},
\end{equation}
\begin{equation}\label{12}
\mathbf{U}_{n,n} = 1, \quad 1\leq n\leq N+1.
\end{equation}
In order to deal with probability operation, we introduce the auxiliary variable $t_{i}$ and convert the \eqref{8e} into
\begin{align}\label{13}
\mathbb{P}\Bigg(\frac{|\mathbf{h}_{\text{I},\text{E}}^{H}\bm{\Theta}\mathbf{h}_{\text{B},\text{I}}+\text{h}_{\text{B},\text{E}}^{H}|^{2}P_{i}}
{\sum_{j=i+1}^{K}|\mathbf{h}_{\text{I},\text{E}}^{H}\bm{\Theta}\mathbf{h}_{\text{B},\text{I}}+\text{h}_{\text{B},\text{E}}^{H}|^{2}P_{j}+\sigma^{2}}
>t_{i}\Bigg) \leq p_{\text{max},i},\quad 1\leq i\leq K,
\end{align}
where $t_{i}$ satisfies
\begin{equation}\label{14}
R_{\text{s},i}\geq\log_{2}\left(1+\frac{
(\text{Tr}(\mathbf{J}_{i}\mathbf{U})+|h_{\text{B},i}^{H}|^{2})P_{i}}{\sum_{j=i+1}^{K}(\text{Tr}(\mathbf{J}_{i}\mathbf{U})+|h_{\text{B},i}^{H}|^{2})
P_{j}+\sigma^{2}}\right)-\log_{2}(1+t_{i}),\quad 1\leq i\leq K.
\end{equation}
Then, by applying Proposition 1, we transform the probabilistic constraint \eqref{13} into a deterministic form, as shown below.
\begin{proposition}
    For the independent Rayleigh fading channels $\mathbf{h}_{\text{I}_{l},\text{E}}\sim \mathcal{CN}(0,L_{\text{I}_{l},\text{E}}^{2}\mathbf{I})$ and $\text{h}_{\text{B},\text{E}}\sim \mathcal{CN}(0,L_{\text{B},\text{E}}^{2})$, the SOP constraint \eqref{13} can be rewritten as
    \begin{equation}\label{15}
    t_{i}\geq\frac{\log\left(\frac{1}{p_{\text{max},i}}\right)(\xi_{\text{E}}^{2}+|L_{\text{B},\text{E}}|^{2})P_{i}}
    {\log\left(\frac{1}{p_{\text{max},i}}\right)\sum_{j=i+1}^{K}(\xi_{\text{E}}^{2}+|L_{\text{B},\text{E}}|^{2})P_{j}+\sigma^{2}},\quad 1\leq i\leq K,
    \end{equation}
    where $\xi_{\text{E}}^{2}=\sum_{l=1}^{L}|L_{\text{B},\text{I}_{l}}L_{\text{I}_{l},\text{E}}|^{2}N_{l}$, while $L_{\text{I}_{l},\text{E}}$, $L_{\text{B},\text{I}_{l}}$ and $L_{\text{B},\text{E}}$ denote the large-scale path losses of $\text{IRS}_{l}$-E, BS-$\text{IRS}_{l}$ and BS-E links, respectively.
\end{proposition}
\begin{IEEEproof}
See Appendix A.
\end{IEEEproof}

Exploiting Proposition 1 and substituting \eqref{14} into objective function, problem (8) is transformed to
\begin{subequations}
\begin{align}
&\max\limits_{\mathbf{U},\mathbf{\bar{u}},P_{i}}\quad \min\limits_{1\leq i\leq K}\  \log_{2}\left(1+\frac{(\text{Tr}(\mathbf{J}_{i}\mathbf{U})+|h_{\text{B},i}^{H}|^{2})P_{i}}{\sum_{j=i+1}^{K}(\text{Tr}(\mathbf{J}_{i}\mathbf{U})+|h_{\text{B},i}^{H}|^{2})
P_{j}+\sigma^{2}}\right)-\log_{2}\Big(1+\underline{t}_{i}\Big),\\
\label{16b}&\quad\text{s.t.} \quad\quad \eqref{8b}, \eqref{9}, \eqref{11}, \eqref{12},
\end{align}
\end{subequations}
where $\underline{t}_{i}=\frac{\log\left(\frac{1}{p_{\text{max},i}}\right)(\xi_{\text{E}}^{2}+|L_{\text{B},\text{E}}|^{2})P_{i}}
    {\log\left(\frac{1}{p_{\text{max},i}}\right)\sum_{j=i+1}^{K}(\xi_{\text{E}}^{2}+|L_{\text{B},\text{E}}|^{2})P_{j}+\sigma^{2}}$. Note that problem (16) is equivalent to problem (8) because the lower bound of $R_{\text{s},{i}}$ in \eqref{14} is a monotone increasing function of $t_{i}$, and the constraint \eqref{15} is active when the objective reaches maximum value. However, problem (16) is still non-convex owing to the coupled variables and the rank-one constraint \eqref{9}.
To decompose the coupled variables, we introduce auxiliary variables $g_{i}$, $\upsilon_{i}$, $\upsilon_{\text{E},i}$ and $\upsilon_{\text{min},i}$, which satisfy

\begin{subequations}
\begin{gather}
\label{17a}g_{i}\left(\text{Tr}(\mathbf{J}_{i}\mathbf{U})+|h_{\text{B},i}^{H}|^{2}\right)\geq1,\quad 1\leq i\leq K,\\
\label{17b}1+\frac{P_{i}}{\sum_{j=i+1}^{K}P_{j}+\sigma^{2}g_{i}}\geq \upsilon_{i},\\
\label{17c} 1+\frac{\log\left(\frac{1}{p_{\text{max},i}}\right)(\xi_{\text{E}}^{2}+|L_{\text{B},\text{E}}|^{2})P_{i}}
    {\log\left(\frac{1}{p_{\text{max},i}}\right)\sum_{j=i+1}^{K}(\xi_{\text{E}}^{2}+|L_{\text{B},\text{E}}|^{2})P_{j}+\sigma^{2}}\leq \upsilon_{\text{E},i},\\
\label{17d}\frac{\upsilon_{i}}{\upsilon_{\text{E},i}}\geq \upsilon_{\text{min},i}.
\end{gather}
\end{subequations}
For \eqref{17a}, we directly rewrite it as the form of convex linear matrix inequality (LMI):
\begin{equation}
\label{18}
\begin{bmatrix}
g_{i} & 1 \\
1 & \text{Tr}(\mathbf{J}_{i}\mathbf{U})+|h_{\text{B},i}^{H}|^{2}
\end{bmatrix} \succeq \mathbf{0},\quad 1\leq i\leq K.
\end{equation}
While for the inequalities \eqref{17b} and \eqref{17d}, we employ the arithmetic-geometric mean (AGM) inequality \cite{W.Zhang_PLS_2020TVT} and transform them into
\begin{equation}
\label{19}
\left(\left(\upsilon_{i}-1\right)\varpi_{i}\right)^2+\left(\left(\sum\nolimits_{j=i+1}^{K}P_{j}
+\sigma^{2}g_{i}\right)/\varpi_{i}\right)^2 \leq 2P_{i},
\end{equation}
\begin{equation}
\label{20}
\left(\upsilon_{\text{E},i}\varpi_{\text{min},i}\right)^{2}+(\upsilon_{\text{min},
i}/\varpi_{\text{min},i})^{2}\leq 2\upsilon_{i},
\end{equation}
where equalities hold if and only if when $\varpi_{i}=\sqrt{\left(\sum\nolimits_{j=i+1}^{K}P_{j}
+\sigma^{2}g_{i}\right)/\left(\upsilon_{i}-1\right)}$ and $\varpi_{\text{min},i}=\sqrt{\upsilon_{\text{min},i}/\upsilon_{\text{E},i}}$.
Moreover, by introducing slack variable $\omega_{i}$, we transform \eqref{17c} into
\begin{subequations}
\begin{gather}
\label{21a} \log\left(\frac{1}{p_{\text{max},i}}\right)(\xi_{\text{E}}^{2}+|L_{\text{B},\text{E}}|^{2})P_{i}\leq \omega_{i}^{2},\quad 1\leq i\leq K,\\
\label{21b} \omega_{i}^{2} \leq (\upsilon_{\text{E},i}-1)\left(\log\left(\frac{1}{p_{\text{max},i}}\right)\sum_{j=i+1}^{K}(\xi_{\text{E}}^{2}+|L_{\text{B},\text{E}}|^{2})P_{j}+\sigma^{2}\right),\quad 1\leq i\leq K.
\end{gather}
\end{subequations}
Afterwards, the first-order Taylor expansion is  utilized to rewrite \eqref{21a} as
\begin{equation}
\label{22} \log\left(\frac{1}{p_{\text{max},i}}\right)(\xi_{\text{E}}^{2}+|L_{\text{B},\text{E}}|^{2})P_{i} \leq 2\tilde{\omega}_{i}\omega_{i}-\tilde{\omega}^{2}_{i},\quad 1\leq i\leq K,
\end{equation}
where $\tilde{\omega}_{i}$ denotes the given local point generated in the previous iteration. The quadratic constraint \eqref{21b} can be rewritten as
\begin{equation}
\label{23}
\begin{bmatrix}
\upsilon_{\text{E},i}-1 & \omega_{i} \\
\omega_{i} & \log\left(\frac{1}{p_{\text{max},i}}\right)\sum_{j=i+1}^{K}(\xi_{\text{E}}^{2}+|L_{\text{B},\text{E}}|^{2})P_{j}+\sigma^{2}
\end{bmatrix}
\succeq \mathbf{0},\quad 1\leq i\leq K.
\end{equation}


\begin{figure}[h]
\centering
\subfigure[The unit-modulus region of $\mathbf{\bar{u}}_{n}$.]{
\includegraphics[scale = 0.7]{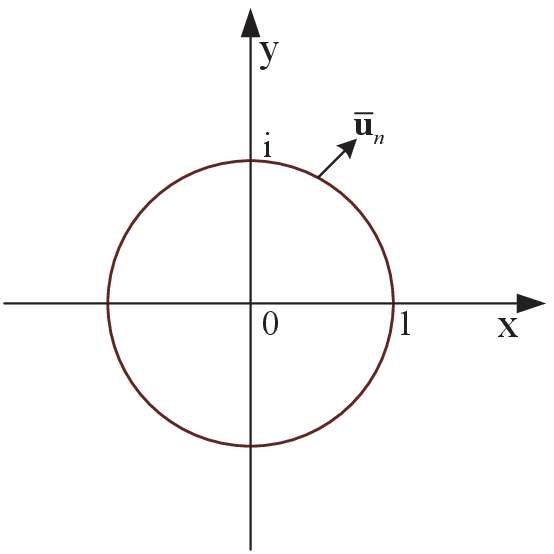}
\label{Fig.2(a)}
}
\quad
\subfigure[The ring-modulus region of $\mathbf{\bar{u}}_{n}$.]{
\includegraphics[scale = 0.7]{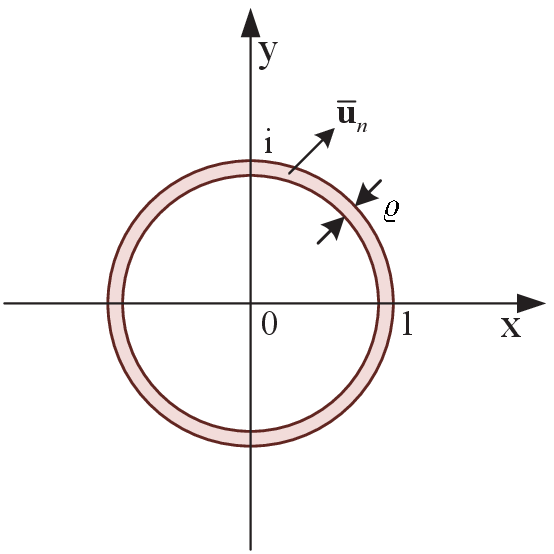}
\label{Fig.2(b)}
}
\caption{Illustration of the ring-penalty method.}
\end{figure}
To handle the non-convex rank-one constraint \eqref{9}, we propose a ring-penalty method in this paper, which relaxes the unit-modulus region of $\mathbf{\bar{u}}_{n}$ into the ring area with width $\varrho$, as depicted in Fig. 2. Mathematically, we first relax \eqref{9} into the convex LMI form, i.e.,
\begin{equation}
\label{24}
\begin{bmatrix}
1 & \mathbf{\bar{u}}^{H} \\
\mathbf{\bar{u}} & \mathbf{U}
\end{bmatrix} \succeq \mathbf{0}.
\end{equation}
Next, to ensure the equivalence between \eqref{24} and \eqref{9}, we have following modulus constraint
\begin{gather}
\label{25}
|\mathbf{\bar{u}}_{n}|^{2}\geq 1-\varrho, \quad 1\leq n\leq N+1,
\end{gather}
where $\varrho>0$ denotes the penalty term. We note that \eqref{24} and \eqref{25} are equivalent to \eqref{9} when $\varrho\rightarrow0$. Nevertheless, since \eqref{25} is non-convex and can not be directly dealt with, we adopt the first-order Taylor expansion to transform it into
\begin{gather}
\label{26}
2\Re(\mathbf{\bar{u}}^{[\text{p}]}_{n}\mathbf{\bar{u}}_{n}^{H})-|\mathbf{\bar{u}}^{[\text{p}]}_{n}|^{2}\geq 1-\varrho, \quad 1\leq n\leq N+1,
\end{gather}
where the left-hand side of \eqref{26} denotes the first-order Taylor approximation of $|\mathbf{\bar{u}}_{n}|^{2}$ at point $\mathbf{\bar{u}}^{[\text{p}]}_{n}$. As such, we reformulate problem (8) as
\begin{subequations}
\begin{align}
\label{27a} &\max\limits_{\mathbf{U},\mathbf{\bar{u}},P_{i},g_{i}, \upsilon_{i}, \upsilon_{\text{E},i}, \upsilon_{\text{min},i},\omega_{i},\varrho}
\quad \min\limits_{1\leq i\leq K}\ \upsilon_{\text{min},i}-\tau\varrho,\\
\label{27b}&\quad\quad\ \text{s.t.} \quad \eqref{8b},\eqref{11},\eqref{18}-\eqref{20},\eqref{22}-\eqref{24},\eqref{26},
\end{align}
\end{subequations}
where $\tau>0$ represents the constant scaling factor for the penalty term $\varrho$. The optimization problem (27) can be efficiently solved by the CVX toolbox, and then, we summarize the overall algorithm in \textbf{Algorithm-1}, where $\epsilon$ and $\varrho_{\text{t}}$ represent the stopping criterion and rank-one accuracy, respectively.

\begin{table}[h]
    \centering
    \begin{tabular}{p{445pt}}
    \toprule
    \textbf{Algorithm-1:} Ring-penalty Based SCA Algorithm \\
    \midrule
    1: \textbf{Initialization}: Initialize the iteration parameters as $\varpi_{\text{min},i}(n)$, $\varpi_{i}(n)$, $\tilde{\omega}_{i}(n)$, $\mathbf{\bar{u}}^{[\text{p}]}(n)$ and $\tau$ with $n=1$;\\
    \hangafter 1 
    \hangindent 2em 
    2: \textbf{Repeat}\\
    3:          \quad Solve the convex problem (27), and obtain the \textit{n}th optimal solutions
                $\mathbf{U}^{*}(n)$, $\mathbf{\bar{u}}^{*}(n)$, $P_{i}^{*}(n)$, $g_{i}^{*}(n)$, $\upsilon_{i}^{*}(n)$, $\upsilon_{\text{E},i}^{*}(n)$, \\
                \qquad $\upsilon_{\text{min},i}^{*}(n)$, $\omega_{i}^{*}(n)$ and $\varrho^{*}(n)$;\\
    4:          \quad Set $n = n+1$;\\
    5:          \quad Update the iteration parameters $\mathbf{\bar{u}}^{[\text{p}]}(n)=\mathbf{\bar{u}}^{*}(n-1)$,  $\varpi_{\text{min},i}(n)=\sqrt{\upsilon_{\text{min},i}^{*}(n-1)/\upsilon_{\text{E},i}^{*}(n-1)}$, $\varpi_{i}(n)=$\\
    \qquad $\sqrt{\left(\sum\nolimits_{j=i+1}^{K}P_{j}^{*}(n-1)
        +\sigma^{2}g_{i}^{*}(n-1)\right)/\left(\upsilon_{i}^{*}(n-1)-1\right)}$ and $\tilde{\omega}_{i}(n)=\omega_{i}^{*}(n-1)$;\\
    6:\textbf{Until} $|\upsilon_{\text{min},i}^{*}(n)-\upsilon_{\text{min},i}^{*}(n-1)|\leq\epsilon$ and $\varrho^{*}(n)\leq\varrho_{\text{t}}$.\\
    \bottomrule
    \end{tabular}
\end{table}

\subsection{Convergence and Complexity Analysis}
Note that the proposed ring-penalty based SCA algorithm is guaranteed to converge with the non-increasing objective value over iterations. Specifically, we can denote the objective value as the function of the optimization variable set $\mathcal{X}=\{\mathbf{U},\mathbf{\bar{u}},P_{i},g_{i}, \upsilon_{i}, \upsilon_{\text{E},i}, \upsilon_{\text{min},i},\omega_{i},\varrho\}$, i.e., $g(\mathcal{X})$. According to \cite[Lemma 2.2]{SCA}, the sequence $\{g(\mathcal{X})\}$ generated by the SCA iterations remains non-decreasing over the compact and non-empty feasible set, i.e.,

\begin{equation}
\label{32}
g\left(\mathcal{X}(n)\right)\leq g\left(\mathcal{X}(n+1)\right).
\end{equation}

Thus, $\{g(\mathcal{X})\}$ is bounded by the limited value, which guarantees the convergence of the SCA algorithm in \textbf{Algorithm-1}.

Moreover, as for the optimization problem (27), it includes $1$ LMI constraint of dimension $N+2$ and $2K$ LMI constraint of dimension $2$, $2K+N+1$ linear constraints, and $2K$ second-order cone constraints of dimension $3$. The generic interior-point method can be employed to solve it with the computational complexity
$\mathcal{O}\Big(l_{\text{s}}\log(1/\epsilon)\sqrt{\Delta} \{d_{n}[(N+2)^3+36K+N+1]+d_{n}^{2}[(N+2)^{2}+10K+N+1]+d_{n}^{3}\}\Big)$ \cite{W.Zhang_PLS_2020TVT}, where $l_{\text{s}}$ denotes the number of iterations, the barrier parameter satisfies $\Delta=10K+2N+3$ \cite{K.Wang_Complexity-1}, and the number of decision variables $d_{n}$ equals to $(N+1)^{2}+N+6K+2$.

\section{General Multi-Antenna System}\label{multi_antenna_case}

In this section, we address the general case that the BS is equipped with multiple antennas. Unlike the SISO NOMA networks, the MISO NOMA network enabling beamforming structure is rather challenging to design since channel order $\|\mathbf{h}_{\text{I},i}^{H}\bm{\Theta}\mathbf{H}_{\text{B},\text{I}}
+\mathbf{h}_{\text{B},i}^{H}\|^{2}>\|\mathbf{h}_{\text{I},j}^{H}\bm{\Theta}\mathbf{H}_{\text{B},\text{I}}
+\mathbf{h}_{\text{B},j}^{H}\|^{2}$ does not necessarily lead to $R_{i,j}> R_{j,j}$ \cite{Y.Liu_MIMO_NOMA}, which is difficult to handle as beamforming vectors and reflection coefficients are highly coupled. To tackle the challenging issue, an efficient AO algorithm is proposed in this post, which divides the original problem into the two subproblems and optimize the transmit beamforming and reflection coefficients alternately.


\subsection{Transmit Beamforming Optimization}\label{sec:minimize_power_pf}
By defining $\mathbf{h}_{i}^{H}=\mathbf{h}_{\text{I},i}^{H}\bm{\Theta}\mathbf{H}_{\text{B},\text{I}}+\mathbf{h}_{\text{B},i}^{H}$, $\mathbf{H}_{i}=\mathbf{h}_{i}\mathbf{h}_{i}^{H}$, and $\mathbf{W}_{i}=\mathbf{w}_{i}\mathbf{w}_{i}^{H}$, the terms of $|(\mathbf{h}_{\text{I},i}^{H}\bm{\Theta}\mathbf{H}_{\text{B},\text{I}}+\mathbf{h}_{\text{B},i}^{H})\mathbf{w}_{i}|^{2}$ can be written as $\text{Tr}(\mathbf{H}_{i}\mathbf{W}_{i})$ for $1\leq i\leq K$. Thus, with the fixed reflection coefficients $\bm{\Theta}$, the original problem (7) is reduced to

\begin{subequations}
\begin{align}
\label{29a} &\max\limits_{\mathbf{W}_{i},\mathbf{w}_{i},R_{\text{s},{i}}}\quad \min\limits_{1\leq i\leq K}\  R_{\text{s},{i}},\\
\label{29b}&\quad\text{s.t.} \quad \sum\nolimits_{i=1}^{K}\text{Tr}(\mathbf{W}_{i})\leq P_{\text{B}},\\
\label{29c}&\quad\quad\quad\,\ \text{Tr}(\mathbf{H}_{i}\mathbf{W}_{i})\leq\dots\leq\text{Tr}(\mathbf{H}_{i}\mathbf{W}_{1}),\quad 1\leq i\leq K,\\
\label{29d}&\quad\quad\quad\,\, R_{i,i}\leq R_{k,i}, \quad 1\leq i\leq k\leq K,\\
\label{29e}&\quad\quad\quad\,\, \mathbb{P}(R_{\text{E},i}>R_{i,i}-R_{\text{s},i}) \leq p_{\text{max},i},\quad 1\leq i\leq K,\\
\label{29f}&\quad\quad\quad\,\, \mathbf{W}_{i}=\mathbf{w}_{i}\mathbf{w}_{i}^{H},\quad 1\leq i\leq K,
\end{align}
\end{subequations}

where $R_{k,i}=\log_{2}\Big(1+\frac{\text{Tr}(\mathbf{H}_{k}\mathbf{W}_{i})}{\sum_{j=i+1}^{K}\text{Tr}(\mathbf{H}_{k}\mathbf{W}_{j})+\sigma^{2}}\Big)$, while the eavesdropping rate is equivalently represented as $R_{\text{E},i}=\log_{2}\left(1+\frac{(\mathbf{h}_{\text{I},\text{E}}^{H}\bm{\Theta}\mathbf{H}_{\text{B},\text{I}}+\mathbf{h}_{\text{B},\text{E}}^{H})\mathbf{W}_{i}
(\mathbf{H}_{\text{B},\text{I}}^{H}\bm{\Theta}^{H}\mathbf{h}_{\text{I},\text{E}}+\mathbf{h}_{\text{B},\text{E}})}{\sum_{j=i+1}^{K}
(\mathbf{h}_{\text{I},\text{E}}^{H}\bm{\Theta}\mathbf{H}_{\text{B},\text{I}}+\mathbf{h}_{\text{B},\text{E}}^{H})\mathbf{W}_{j}
(\mathbf{H}_{\text{B},\text{I}}^{H}\bm{\Theta}^{H}\mathbf{h}_{\text{I},\text{E}}+\mathbf{h}_{\text{B},\text{E}})+\sigma^{2}}\right)$. Note that problem (29) is intractable to solve since constraints \eqref{29d}-\eqref{29f} are non-convex. In order to tackle the problem (29), some reasonable transformations and safe approximations will be adopted in the following, which convert (29) into the convex programming that can be directly solved by the CVX toolbox.

To start with, we introduce a slack variable $z_{k,i}$ for $1\leq i\leq k\leq K$, which satisfies
\begin{equation}
\label{30}
z_{k,i}\leq\frac{\text{Tr}(\mathbf{H}_{k}\mathbf{W}_{i})}{\sum_{j=i+1}^{K}\text{Tr}(\mathbf{H}_{k}\mathbf{W}_{j})+\sigma^{2}}.
\end{equation}
Similar to \eqref{19}, we apply the AGM inequality to rewrite \eqref{30} as the convex form, i.e.,
\begin{equation}
\label{31}
(z_{k,i}\varpi_{k,i})^{2}+\left(\frac{\sum_{j=i+1}^{K}\text{Tr}(\mathbf{H}_{k}\mathbf{W}_{j})+\sigma^{2}}{\varpi_{k,i}}\right)^{2}\leq2\text{Tr}(\mathbf{H}_{k}\mathbf{W}_{i}),
\end{equation}
where the equality holds if and only if when $\varpi_{k,i}=\sqrt{\frac{\sum_{j=i+1}^{K}\text{Tr}(\mathbf{H}_{k}\mathbf{W}_{j})+\sigma^{2}}{z_{k,i}}}$. Then, exploiting \eqref{31} and Lemma 1, we transform constraint \eqref{29d} into

\begin{equation}
\label{32}
z_{i,i}\leq z_{k,i},\quad 1\leq i\leq k\leq K.
\end{equation}

\begin{lemma}
    When the objective function reaches the optimum, the achievable rate of $\text{U}_{i}$ decoding its own signal equals to $\log_{2}(1+z_{i,i})$, i.e.,
    \begin{equation}
    \label{33}
    z_{i,i}= \frac{\text{Tr}(\mathbf{H}_{i}\mathbf{W}_{i})}{\sum_{j=i+1}^{K}\text{Tr}(\mathbf{H}_{i}\mathbf{W}_{j})+\sigma^{2}},\quad 1\leq i\leq K.
    \end{equation}
\end{lemma}
\begin{IEEEproof}
See Appendix B.
\end{IEEEproof}

Recall that in \eqref{13} and \eqref{14}, the auxiliary variable $t_{i}$ is introduced to simplify the probabilistic constraint \eqref{29e}. The transformed SOP constraint of $\text{U}_{i}$ is given by
\begin{align}\label{34}
\mathbb{P}\Bigg((\mathbf{h}_{\text{I},\text{E}}^{H}\bm{\Theta}\mathbf{H}_{\text{B},\text{I}}+\mathbf{h}_{\text{B},\text{E}}^{H})\mathbf{E}_{i}(\mathbf{H}_{\text{B},\text{I}}^{H}\bm{\Theta}^{H}\mathbf{h}_{\text{I},\text{E}}+\mathbf{h}_{\text{B},\text{E}})
>t_{i}\sigma^{2}\Bigg) \leq p_{\text{max},i},
\end{align}
where $\mathbf{E}_{i}=\mathbf{W}_{i}-t_{i}\sum_{j=i+1}^{K}\mathbf{W}_{j}$, and $t_{i}$ satisfies
\begin{equation}\label{35}
R_{\text{s},i}\geq\log_{2}\Big(1+\frac{\text{Tr}(\mathbf{H}_{i}\mathbf{W}_{i})}{\sum_{j=i+1}^{K}\text{Tr}(\mathbf{H}_{i}\mathbf{W}_{j})+\sigma^{2}}\Big)-\log_{2}(1+t_{i})
\geq\log_{2}\left(\frac{1+z_{i,i}}{1+t_{i}}\right).
\end{equation}
To further convert the probabilistic constraint into the tractable form, a conservative transformation based on Bernstein-type inequality \cite{Bernstein} is introduced as follows.
\begin{proposition}
    With the independent Rayleigh fading channels $\mathbf{h}_{\text{I}_{l},\text{E}}\sim \mathcal{CN}(0,L_{\text{I}_{l},\text{E}}^{2}\mathbf{I})$ and $\mathbf{h}_{\text{B},\text{E}}\sim \mathcal{CN}(0,L_{\text{B},\text{E}}^{2}\mathbf{I})$, the approximated SOP constraint \eqref{34} can be represented as
    \begin{subequations}
    \begin{gather}
    \label{36a}t_{i}\geq \frac{1}{\sigma^{2}}\left( \text{Tr}(\bm{\Phi}_{i})+\sqrt{2\log\left(\frac{1}{p_{\text{max},i}}\right)}\|\bm{\Phi}_{i}\|_{\text{F}}+\log\left(\frac{1}{p_{\text{max},i}}\right)\phi_{i}\right),\quad 1\leq i\leq K, \\
    \label{36b} \phi_{i}\mathbf{I}-\bm{\Phi}_{i}\succeq \mathbf{0},\quad 1\leq i\leq K,
    \end{gather}
    \end{subequations}
    where $\mathbf{\overline{L}}_{\text{I},\text{E}}=\text{blkdiag}([\mathbf{L}_{\text{I}_{1},\text{E}},\dots,
    \mathbf{L}_{\text{I}_{L},\text{E}}])$ with $\mathbf{L}_{\text{I}_{l},\text{E}}=\text{diag}([L_{\text{I}_{l},\text{E}},\dots,L_{\text{I}_{l},\text{E}}]^{T})\in\mathbb{C}^{N_{l}\times N_{l}}$. The joint beamforming matrix is given by
    \begin{equation}\label{37}
    \bm{\Phi}_{i}=\begin{bmatrix}
    L_{\text{B},\text{E}}^{2}\mathbf{W}_{i} &  L_{\text{B},\text{E}}\mathbf{W}_{i}\mathbf{H}_{\text{B},\text{I}}^{H}\bm{\Theta}^{H}\mathbf{\overline{L}}_{\text{I},\text{E}}\\
    \mathbf{\overline{L}}_{\text{I},\text{E}}\bm{\Theta}\mathbf{H}_{\text{B},\text{I}}\mathbf{W}_{i}L_{\text{B},\text{E}} & \mathbf{\overline{L}}_{\text{I},\text{E}}\bm{\Theta}\mathbf{H}_{\text{B},\text{I}}\mathbf{W}_{i}\mathbf{H}_{\text{B},\text{I}}^{H}\bm{\Theta}^{H}\mathbf{\overline{L}}_{\text{L},\text{E}}
    \end{bmatrix},\quad 1\leq i\leq K.
    \end{equation}
\end{proposition}
\begin{IEEEproof}
See Appendix C.
\end{IEEEproof}
\begin{remark}
    (Difference between Single and Multiple Antenna Systems) To guarantee secure legitimate transmission, the core idea is to enable IRSs to improve channel qualities of NOMA users and degrade channel quality of E. However, it is verified that maximum eavesdropping (signal-to-interference-plus-noise ratio) SINR in single-antenna BS case is upper-bounded by right-hand side of \eqref{15}, which, however, can not be reconfigured by IRSs. Therefore, when IRSs can not provide the greater legitimate channel gains than $\log\left(\frac{1}{p_{\text{max},i}}\right)(\xi_{\text{E}}^{2}+|L_{\text{B},\text{E}}|^{2})$, E can always obtain the confidential messages of users, which limits secrecy performance of single-antenna networks. While for multi-antenna scenario, the upper bound SINR of eavesdropper given in right-hand side of \eqref{36a} is highly affected by IRS phase shifts. This indicates that even without eavesdropper's instantaneous CSI, IRSs are capable of deteriorating the signal reception of the eavesdropper, which demonstrates the secrecy potential of IRS integrating multi-antenna BS networks.
\end{remark}

According to Proposition 2,
As for the non-convex rank-one constraint \eqref{29f}, the proposed ring-penalty method is not applicable since the elements of $\mathbf{w}_{i}$ do not meet the constant-modulus condition. Here, we adopt the DCR method \cite{X.Yu_imperfectCSI,T.Jiang_DCP} to tackle this issue. For exposition purpose, we first rewrite the \eqref{29f} as
\begin{subequations}
    \begin{gather}
    \label{38a}\mathbf{W}_{i}\succeq \mathbf{0}, \\
    \label{38b} \text{rank}(\mathbf{W}_{i})=1.
    \end{gather}
\end{subequations}
According to the \cite{X.Yu_imperfectCSI}, \eqref{38b} is transformed into $\text{Tr}(\mathbf{W}_{i})=\|\mathbf{W}_{i}\|_{2}$, with $\text{Tr}(\mathbf{W}_{i})=\sum\nolimits_{i=1}^{N}\sigma_{i}$ and $\|\mathbf{W}_{i}\|_{2}=\sigma_{1}$. With \cite[Prop. 2]{T.Jiang_DCP}, we further relax the rank-one constraint into the form of the difference-of-convex constraint, which is given by
\begin{equation}
\label{39}
\Re(\text{Tr}(\mathbf{W}_{i}^{H}(\mathbf{I}-\mathbf{w}_{i,1}\mathbf{w}_{i,1}^{H})))\leq\varrho,
\end{equation}
where $\mathbf{w}_{i,1}$ denotes the leading eigenvector of $\mathbf{W}_{i}$ obtained in the previous iteration, and $\varrho$ is the penalty factor.

As a result, problem (29) can be reformulated as
\begin{subequations}
\begin{align}
\label{40a} &\max\limits_{\mathbf{W}_{i},\mathcal{Z},t_{i},\bm{\Phi}_{i},\phi_{i},\varrho}\quad \min\limits_{1\leq i\leq K}\  \frac{1+z_{i,i}}{1+t_{i}}-\tau\varrho,\\
\label{40b}&\quad\quad\text{s.t.} \quad \eqref{29b},\eqref{29c},\eqref{31},\eqref{32},\eqref{36a},\eqref{36b},\eqref{37},\eqref{38a},\eqref{39},
\end{align}
\end{subequations}
where $\mathcal{Z}=\{z_{k,i}|1\leq i\leq k\leq K\}$. Since the problem (41) is a typical fractional programming, we employ the Dinkelbach algorithm \cite{Dinkelbach} to optimally solve it, which transforms the (40) into the following parametric form
\begin{subequations}
\begin{align}
\label{41a} &\max\limits_{\mathbf{W}_{i},\mathcal{Z},t_{i},\bm{\Phi}_{i},\phi_{i},\varrho,\zeta}\quad \  \zeta-\tau\varrho,\\
\label{41b}&\quad\quad\text{s.t.} \quad 1+z_{i,i}-\mu_{i}(1+t_{i})\geq\zeta,\quad 1\leq i\leq K,\\
\label{41c}&\quad\quad\quad\quad \eqref{29b},\eqref{29c},\eqref{31},\eqref{32},\eqref{36a},\eqref{36b},\eqref{37},\eqref{38a},\eqref{39},
\end{align}
\end{subequations}
where $\mu_{i}$ starts from $0$, and is updated by $\mu_{i}=\frac{1+z_{i,i}}{1+t_{i}}$ at each iteration, while the auxiliary variable $\zeta$ is introduced to measure the approximation gap between $\mu_{i}$ and the term of $\frac{1+z_{i,i}}{1+t_{i}}$. Note that (41) is a convex optimization problem and can be efficiently solved  by the convex solver CVX in an iterative manner.

\subsection{Reflection Coefficients Optimization}\label{sec:minimize_power_pf}

By fixing the transmit beamforming $\mathbf{W}_{i}$ and defining $\mathbf{G}_{i}=[\text{diag}(\mathbf{h}_{\text{I},i}^{H})\mathbf{H}_{\text{B},\text{I}};\mathbf{h}_{\text{B},i}^{H}]$, $\mathbf{\bar{u}}=[\mathbf{u};1]$, $\mathbf{u}=[e^{j\alpha_{1}},\dots,e^{j\alpha_{N}}]^{H}$ and $\mathbf{U}=\mathbf{\bar{u}}\mathbf{\bar{u}}^{H}$, the problem (7) is reduced to
\begin{subequations}
\begin{align}
\label{42a} &\max\limits_{\mathbf{U},\mathbf{\bar{u}},R_{\text{s},{i}}}\quad \min\limits_{1\leq i\leq K}\  R_{\text{s},{i}},\\
\label{42b}&\quad\text{s.t.} \quad \text{Tr}(\mathbf{G}_{i}\mathbf{W}_{i}\mathbf{G}_{i}^{H}\mathbf{U})\leq\dots\leq\text{Tr}(\mathbf{G}_{i}\mathbf{W}_{1}\mathbf{G}_{i}^{H}\mathbf{U}),\quad 1\leq i\leq K,\\
\label{42c}&\quad\quad\quad\,\, R_{i,i}\leq R_{k,i}, \quad 1\leq i\leq k\leq K,\\
\label{42d}&\quad\quad\quad\,\, \mathbb{P}(R_{\text{E},i}>R_{i,i}-R_{\text{s},i}) \leq p_{\text{max},i},\quad 1\leq i\leq K,\\
\label{42e}&\quad\quad\quad\,\, \mathbf{U}=\mathbf{\bar{u}}\mathbf{\bar{u}}^{H},\\
\label{42f}&\quad\quad\quad\,\, \mathbf{U}_{n,n} = 1,\quad 1\leq n\leq N+1,
\end{align}
\end{subequations}
where $R_{k,i}=\log_{2}\left(1+\frac{\text{Tr}(\mathbf{G}_{k}\mathbf{W}_{i}\mathbf{G}_{k}^{H}\mathbf{U})}{\sum_{j=i+1}^{K}\text{Tr}
(\mathbf{G}_{k}\mathbf{W}_{i}\mathbf{G}_{k}^{H}\mathbf{U})+\sigma^{2}}\right)$, and the wiretapping rate is given by $R_{\text{E},i}=\log_{2}\left(1+\frac{(\mathbf{h}_{\text{I},\text{E}}^{H}\text{diag}(\mathbf{\bar{u}})\mathbf{H}_{\text{B},\text{I}}+\mathbf{h}_{\text{B},\text{E}}^{H})\mathbf{W}_{i}
(\mathbf{H}_{\text{B},\text{I}}^{H}\text{diag}(\mathbf{\bar{u}})^{H}\mathbf{h}_{\text{I},\text{E}}+\mathbf{h}_{\text{B},\text{E}})}{\sum_{j=i+1}^{K}
(\mathbf{h}_{\text{I},\text{E}}^{H}\text{diag}(\mathbf{\bar{u}})\mathbf{H}_{\text{B},\text{I}}+\mathbf{h}_{\text{B},\text{E}}^{H})\mathbf{W}_{j}
(\mathbf{H}_{\text{B},\text{I}}^{H}\text{diag}(\mathbf{\bar{u}})^{H}\mathbf{h}_{\text{I},\text{E}}+\mathbf{h}_{\text{B},\text{E}})+\sigma^{2}}\right)$. Nevertheless, the non-convex constraints \eqref{42c} and \eqref{42f} result in much difficulty to solve \eqref{42a}. Recall \eqref{30}-\eqref{32}, we rewrite the \eqref{42c} as
\begin{subequations}
\begin{gather}
\label{43a} (z_{k,i}\varpi_{k,i})^{2}+\left(\frac{\sum_{j=i+1}^{K}\text{Tr}(\mathbf{G}_{k}\mathbf{W}_{j}\mathbf{G}_{k}^{H}\mathbf{U})+\sigma^{2}}{\varpi_{k,i}}\right)^{2}\leq2\text{Tr}(\mathbf{G}_{k}\mathbf{W}_{i}\mathbf{G}_{k}^{H}\mathbf{U}),\\
\label{43b} z_{i,i}\leq z_{k,i},\quad 1\leq i\leq k\leq K,
\end{gather}
\end{subequations}
in which the equality holds if and only if when $\varpi_{k,i}=\sqrt{\frac{\sum_{j=i+1}^{K}\text{Tr}(\mathbf{G}_{k}\mathbf{W}_{j}\mathbf{G}_{k}^{H}\mathbf{U})+\sigma^{2}}{z_{k,i}}}$.
With the transformations \eqref{34}, \eqref{35} and Proposition 2, we rewrite the SOP constraint of $\text{U}_{i}$ as
\begin{subequations}
\begin{gather}
\label{44a} R_{\text{s},i}\geq\log_{2}\left(\frac{1+z_{i,i}}{1+t_{i}}\right),\\
\label{44b} t_{i}\geq \frac{1}{\sigma^{2}}\left( \text{Tr}(\bm{\Phi}_{i})+\sqrt{2\log\left(\frac{1}{p_{\text{max},i}}\right)}\|\bm{\Phi}_{i}\|_{\text{F}}+\log\left(\frac{1}{p_{\text{max},i}}\right)\phi_{i}\right),\\
\label{44c} \phi_{i}\mathbf{I}-\bm{\Phi}_{i}\succeq \mathbf{0},\\
\label{44d} \bm{\Phi}_{i}=\begin{bmatrix}
    L_{\text{B},\text{E}}^{2}\mathbf{W}_{i} &  L_{\text{B},\text{E}}\mathbf{W}_{i}\mathbf{H}_{\text{B},\text{I}}^{H}\mathbf{\overline{L}}_{\text{I},\text{E}}\text{diag}(\mathbf{\bar{u}})^{H}\\
    \text{diag}(\mathbf{\bar{u}})\mathbf{L}_{\text{I},\text{E}}\mathbf{H}_{\text{B},\text{I}}\mathbf{W}_{i}L_{\text{B},\text{E}} & \text{diag}(\mathbf{\bar{u}})\mathbf{\overline{L}}_{\text{I},\text{E}}\mathbf{H}_{\text{B},\text{I}}\mathbf{W}_{i}\mathbf{H}_{\text{B},\text{I}}^{H}\mathbf{\overline{L}}_{\text{I},\text{E}}\text{diag}(\mathbf{\bar{u}})^{H}
    \end{bmatrix}.
\end{gather}
\end{subequations}
Here, in order to convert \eqref{44d} into the convex LMI form, we use the singular value decomposition (SVD) to represent the constant matrix $\mathbf{\overline{L}}_{\text{I},\text{E}}\mathbf{H}_{\text{B},\text{I}}\mathbf{W}_{i}\mathbf{H}_{\text{B},\text{I}}^{H}\mathbf{\overline{L}}_{\text{I},\text{E}}$ equivalently as $\sum_{q}\mathbf{s}_{i,q}\mathbf{d}_{i,q}$. To proceed, with the definition of $\mathbf{S}_{i,q}=\begin{bmatrix}\text{diag}(\mathbf{s}_{i,q}),\mathbf{0}\end{bmatrix}$ and $\mathbf{D}_{i,q}=\begin{bmatrix} \text{diag}(\mathbf{d}_{i,q}), \mathbf{0}\end{bmatrix}^{T}$, the coupled term in \eqref{44d} can be expressed as 
\begin{align}\nonumber
\text{diag}(\mathbf{\bar{u}})\mathbf{\overline{L}}_{\text{I},\text{E}}\mathbf{H}_{\text{B},\text{I}}\mathbf{W}_{i}\mathbf{H}_{\text{B},\text{I}}^{H}\mathbf{\overline{L}}_{\text{I},\text{E}}\text{diag}(\mathbf{\bar{u}})^{H}&=
\sum_{q}\text{diag}(\mathbf{s}_{i,q})\mathbf{u}\mathbf{u}^{H}\text{diag}(\mathbf{d}_{i,q})\\ \label{45}
&=\sum_{q}\mathbf{S}_{i,q}\mathbf{\bar{u}}\mathbf{\bar{u}}^{H}\mathbf{D}_{i,q}=\sum_{q}\mathbf{S}_{i,q}\mathbf{U}\mathbf{D}_{i,q}.
\end{align}

As such, \eqref{44d} can be reformulated as
\begin{equation}\label{46}
\bm{\Phi}_{i}=\begin{bmatrix}
    L_{\text{B},\text{E}}^{2}\mathbf{W}_{i} &  L_{\text{B},\text{E}}\mathbf{W}_{i}\mathbf{H}_{\text{B},\text{I}}^{H}\mathbf{\overline{L}}_{\text{I},\text{E}}\text{diag}(\mathbf{\bar{u}})^{H}\\
    \text{diag}(\mathbf{\bar{u}})\mathbf{\overline{L}}_{\text{I},\text{E}}\mathbf{H}_{\text{B},\text{I}}\mathbf{W}_{i}L_{\text{B},\text{E}} & \sum_{q}\mathbf{S}_{i,q}\mathbf{U}\mathbf{D}_{i,q}
    \end{bmatrix},\quad(1\leq i\leq K).
\end{equation}
Furthermore, due to the fact that \eqref{42e} and \eqref{42f} are the same as the rank-one constraints \eqref{9} and \eqref{12}, we can relax them as the same forms as \eqref{24} and \eqref{26}. Therefore, the problem (46) is reformulated as
\begin{subequations}
\begin{align}
\label{47a} &\max\limits_{\mathbf{U},\mathbf{\bar{u}},\mathcal{Z},\bm{\Phi}_{i},\phi_{i},\varrho,\zeta}\quad \zeta-\tau\varrho,\\
\label{47b}&\quad\quad\text{s.t.} \quad \frac{1+z_{i,i}}{1+t_{i}}\geq\zeta, \quad(1\leq i\leq K),\\
\label{47c}&\quad\quad\quad\quad \eqref{24},\eqref{26},\eqref{42b},\eqref{42f},\eqref{43a},\eqref{43b},\eqref{44b},\eqref{44c},\eqref{46}.
\end{align}
\end{subequations}
Obviously, the problem (48) is a convex optimization problem  and can be efficiently solved by convex solver.
\begin{remark} (Conservatism of Approximation)
Note that the Berstein-type inequality provides a very conservative approximation for the original SOP constraint when we choose $x=\log\left(\frac{1}{p_{\text{max}}}\right)$ in \eqref{A3-2}. To show the tightness of approximation in Proposition 2, we illustrate the relationship between the actual SOP\footnote{The actual SOP of $\text{U}_{i}$ can be calculated by
$\sum_{i=1}^{N_{\text{p}}}\prod_{j\neq i}^{N_{\text{a}}}\frac{e^{-\frac{\text{Tr}(\bm{\Phi}_{i})+\sqrt{2\log(\frac{1}{p_{\text{max}}})}\|\bm{\Phi}_{i}\|_{\text{F}}+\log(\frac{1}{p_{\text{max}}})
\text{max}\{\rho_{\text{max}}(\bm{\Phi}_{i}),0\}}{\rho_{i}}}}{1-\rho_{j}/\rho_{i}}$ according to \cite{H.-M.Wang_NOMA_PLS}, where $\rho_{j}$ ($1\leq j\leq N_{\text{a}}$) denotes the eigenvalues of $\bm{\Phi}_{i}$, while $\rho_{i}$ ($1\leq i\leq N_{\text{p}}$) denotes the positive eigenvalues of $\bm{\Phi}_{i}$.} and the presupposed SOP $p_{\text{max}}$ in Fig. \ref{Fig.3}. As can be observed, with $p_{\text{max}}$ varying from $0.1$ to $0.9$, the actual SOP is always less than the presupposed SOP, which demonstrates the effectiveness of the proposed AO algorithm since it is capable of ensuring the much lower SOP.

\begin{figure}[h]
  \centering
  \includegraphics[scale = 0.4]{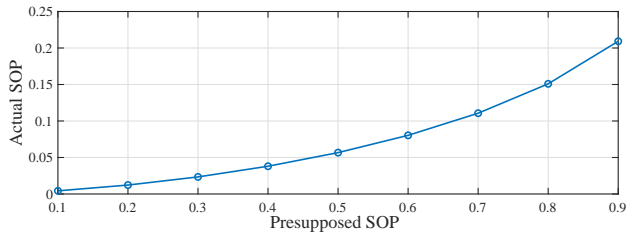}
  \caption{Tightness of the Berstein-type inequality approximation.}
  \label{Fig.3}
\end{figure}
\end{remark}

\subsection{Overall Algorithm, Convergence and Complexity Analysis}\label{Overall}

In the proposed AO algorithm, we optimize the transmit beamforming and reflection coefficients alternately. In detail, a DCR based Dinkelbach algorithm is proposed to optimize the transmit beamforming with the fixed $\bm{\Theta}$, while the globally optimal reflection coefficients is obtained via SCA iterations. More details of the AO algorithm are summarized \textbf{Algorithm-2}.
\begin{table}[h]
    \centering
    \begin{tabular}{p{445pt}}
    \toprule
    \textbf{Algorithm-2:} AO Algorithm \\
    \midrule
    1: \textbf{Initialization}: Initialize $l=1$, $n=1$, $\bm{\Theta}(l-1)$, $\varpi_{k,i}(n)$, $\tau(n)$, $\mu_{i}(n)$, $R_{\text{min},\text{U}}(l-2)=+\infty$ and $R_{\text{min},\text{U}}(l-1)=-\infty$;\\
    \hangafter 1 
    \hangindent 2em 
    2: \textbf{While} $|R_{\text{min}}(l-1)-R_{\text{min}}(l-2)|\geq\epsilon$\\
    3: \ \ \textbf{Repeat}\\
    4: \ \ \quad With the given $\bm{\Theta}(l-1)$, solve the problem (45) and obtain the optimal solutions $\mathbf{W}_{i}^{*}(n)$, $z_{k,i}^{*}(n)$, $t_{i}^{*}(n)$, $\bm{\Phi}_{i}^{*}(n)$, \\
    \ \ \qquad $\phi_{i}^{*}(n)$, $\varrho^{*}(n)$ and $\zeta^{*}(n)$ with $1\leq i \leq k\leq K$; \\
    5: \ \ \quad Set $n = n+1$; \\
    6: \ \ \quad Update the iteration parameters $\varpi_{k,i}(n)=\sqrt{\frac{\sum_{j=i+1}^{K}\text{Tr}(\mathbf{H}_{k}\mathbf{W}_{j}^{*}(n-1))+\sigma^{2}}{z_{k,i}^{*}(n-1)}}$ and $\mu_{i}(n)=\frac{1+z_{i,i}^{*}(n-1)}{1+t_{i}^{*}(n-1)}$;\\
    7: \ \ \textbf{Until} $|\zeta^{*}(n)-\zeta^{*}(n-1)|\leq\epsilon$ and $\varrho^{*}(n)\leq\varrho_{\text{t}}$, output $\mathbf{W}_{i}^{*}(l)$;\\
    8: \ \ Calculate $R_{\text{min},\text{W}}(l)=\log_{2}(\zeta^{*}(n))$ and set $n=1$;\\
    9: \ \ \textbf{Repeat}\\
    10: \quad With the given $\mathbf{W}_{i}(l)$, solve the problem (52) and obtain the optimal solutions $\mathbf{U}^{*}(n)$, $\mathbf{\bar{u}}^{*}(n)$, $z_{k,i}^{*}(n)$, $t_{i}^{*}(n)$, \\
    \ \ \qquad $\bm{\Phi}_{i}^{*}(n)$, $\phi_{i}^{*}(n)$, $\varrho^{*}(n)$ and $\zeta^{*}(n)$ with $1\leq i \leq k\leq K$; \\
    11: \ \ \ Set $n = n+1$; \\
    12: \ \ \  Update the iteration parameters $\varpi_{k,i}(n)=\sqrt{\frac{\sum_{j=i+1}^{K}\text{Tr}(\mathbf{G}_{k}\mathbf{W}_{j}\mathbf{G}_{k}^{H}\mathbf{U}^{*}(n-1))+\sigma^{2}}{z_{k,i}^{*}(n-1)}}$;\\
    13: \  \textbf{Until} $|\zeta^{*}(n)-\zeta^{*}(n-1)|\leq\epsilon$ and $\varrho^{*}(n)\leq\varrho_{\text{t}}$, output $\bm{\Theta}^{*}(l+1)=\text{diag}(\mathbf{u}^{*}(n)[1:N])$ and set $n=1$;\\
    14: \  Calculate $R_{\text{min},\text{U}}(l)=\log_{2}(\zeta^{*}(n))$ and set $l = l+1$; \\
    15: \textbf{End while}.\\
    \bottomrule
    \end{tabular}
\end{table}

The proposed AO algorithm is guaranteed to converge with the non-increasing objective value over iterations. Specifically, we denote the objective value as a function of transmit beamforming and reflection coefficients, e.g., $g(\mathbf{W}_{i},\bm{\Theta})$. In the steps $3$--$8$ of \textbf{Algorithm-2} at the $\textit{l}$th iteration ($l\geq 1$), we perform Dinkelbach iteration to obtain the optimal transmit beamforming $\mathbf{W}_{i}^{*}(l)$ under the given $\bm{\Theta}(l-1)$ of the problem (45). Thus, it follows that $g(\mathbf{W}_{i}(l),\bm{\Theta}(l-1)) \geq g(\mathbf{W}_{i}(l-1),\bm{\Theta}(l-1))$. While in the steps $9$--$14$ of the $\textit{l}$th iteration, we solve the problem (52) to optimize the reflection coefficients with the fixed $\mathbf{W}_{i}(l)$, which leads to $g(\mathbf{W}_{i}(l),\bm{\Theta}(l)) \geq g(\mathbf{W}_{i}(l),\bm{\Theta}(l-1))$. As such, we can obtain the inequality
\begin{gather}
\label{48} g(\mathbf{W}_{i}(l),\bm{\Theta}(l)) \geq g(\mathbf{W}_{i}(l),\bm{\Theta}(l-1)\geq g(\mathbf{W}_{i}(l-1),\bm{\Theta}(l-1)),
\end{gather}
which indicates that the sequence $\{g(\mathbf{W}_{i}(l),\bm{\Theta}(l))\}$ generated by AO algorithm remains non-decreasing over iterations. On the other hand, $g(\mathbf{W}_{i},\bm{\Theta})$ is continuous over the compact feasible set of problem (7) \cite{Convex}, and hence, the upper bound of the objective value is limited by a finite positive number, which thus proves the convergence of the proposed AO algorithm.

Similarly, the whole computational complexity of AO algorithm is given by $\mathcal{O}\Big(l_{\text{AO}}\big(l_{\text{W}}\log(1/\epsilon)\\\sqrt{\Delta_{\text{W}}} \{d_{n,\text{W}}[K(M+N)^3+K(M)^3+(K+1)^{2}+\frac{9}{2}(K^{2}+K)]+d_{n,\text{W}}[K(M+N)^2+K(M)^2+(K+1)^{2}]+d_{n,\text{W}}^{2}\}+l_{\text{U}}\log(1/\epsilon)\sqrt{\Delta_{\text{U}}} \{d_{n,\text{U}}[K(M+N)^3+(N+2)^3+K^{2}+K+2(N+1)+\frac{9}{2}(K^{2}+K)]+d_{n,\text{U}}[K(M+N)^2+(N+2)^2+K^{2}+K+2(N+1)]+d_{n,\text{U}}^{2}\}\big)\Big)$, where $\Delta_{\text{W}}=K(2M+N)+2K^{2}+3K+1$, $\Delta_{\text{U}}=K(M+N)+2K^{2}+2K+3N+4$, $d_{n,\text{W}}=K(M+N)^{2}+M^{2}+\frac{K^{2}+5K}{2}+2$, $d_{n,\text{U}}=(N+1)^{2}+N+\frac{K^{2}+K}{2}+K(M+N)^{2}+3$, $l_{\text{W}}$ and $l_{\text{U}}$ denote the number of iterations for solving problem (41) and (48), while $l_{\text{AO}}$ denotes the number of iteration required for achieving convergence.

\section{Numerical Results}\label{Simulation}
\begin{figure}
  \centering
  \begin{minipage}[b]{0.55\textwidth} 
    \centering
    \includegraphics[width=1\textwidth]{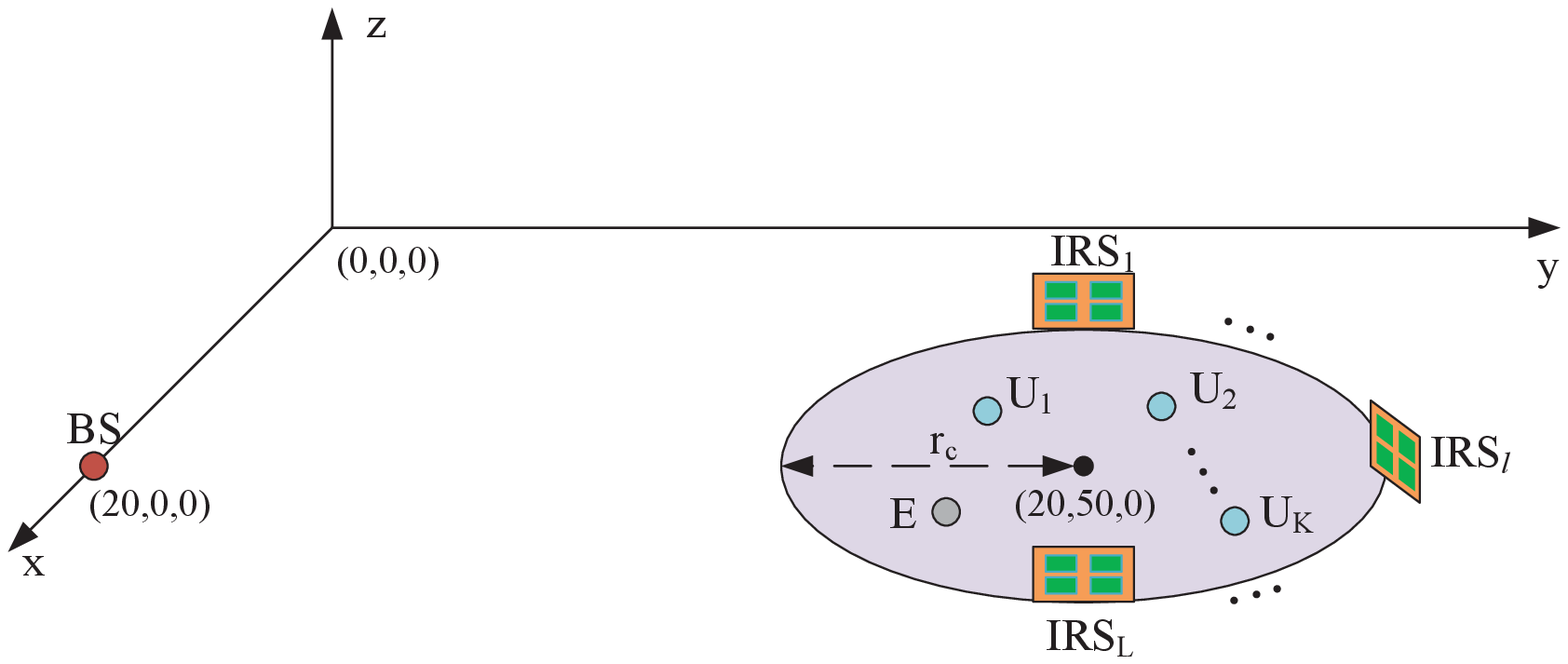}
    \caption{Simulation setup of the considered network.}
    \label{Fig.4}
    \end{minipage}
      \begin{minipage}[b]{0.32\textwidth} 
    \centering 
    \includegraphics[width=1\textwidth]{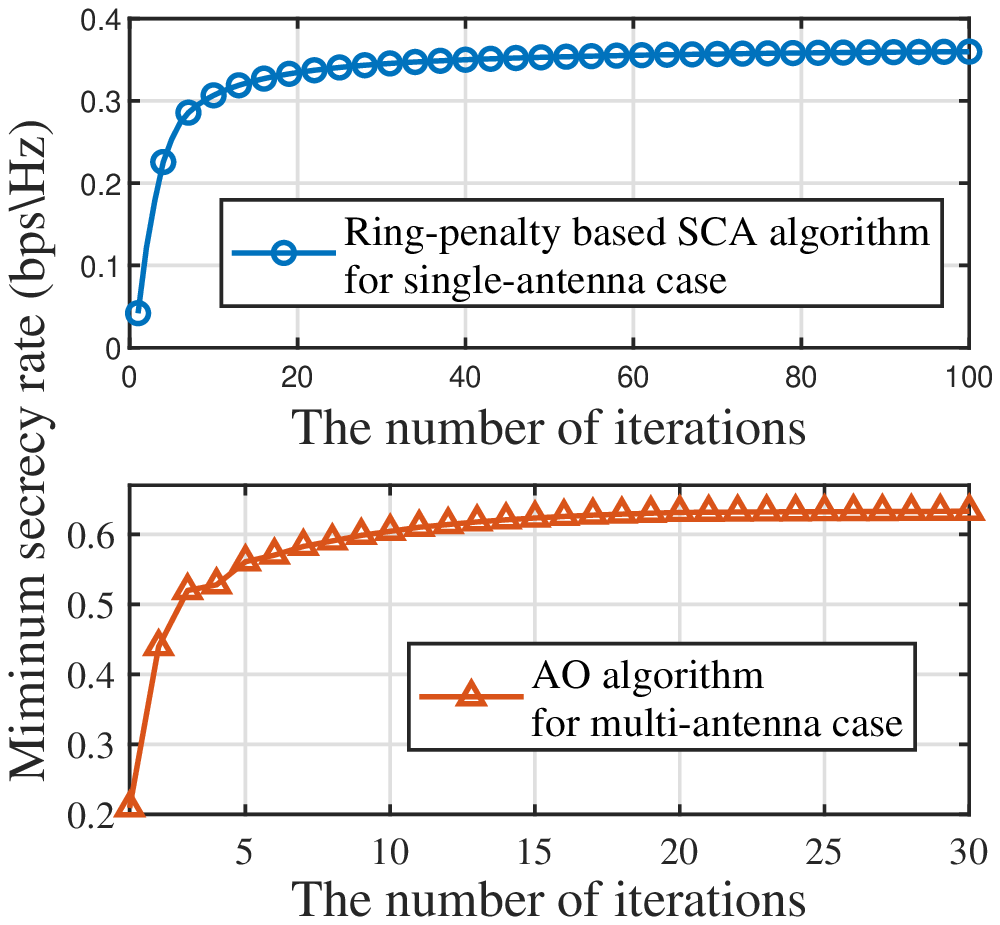} 
    \caption{Convergence of two proposed algorithms for $K=2$, $N=20$, $P_{\text{B}}=15$dBm, $\text{r}_\text{c}=10$m, $p_{\text{max}}=0.1$ and $M=6$.}
    \label{Fig.5}
    \end{minipage}
\end{figure}
In this section, the simulation results are presented to validate the performance of the proposed algorithms. We concentrate on a three-dimensional (3D) coordinate network as shown in Fig. \ref{Fig.4}, where the BS is located at ($20$, $0$, $0$) meter (m), while the E and legitimate users are randomly distributed in the circle centered at ($20$, $50$, $0$) m. For convenience, we assume that all the legitimate users possess the same security requirement, i.e., $p_{\text{max},1}=\dots=p_{\text{max},K}=p_{\text{max}}$. Meanwhile, $L$ IRSs are uniformly distributed on the right half of the circle, with $\text{IRS}_{l}$ being equipped with $N_{l}$ reflecting elements for $1\leq l\leq L$ and $1\leq L$. If not specified, we consider two equivalent IRSs deployment as the distributed scheme, i.e., $L=2$ and $N_{1}=N_{2}=\frac{N}{2}$. Each IRS is equipped with a uniform planar array (UPA) with a half-wavelength antenna spacing. The other simulation parameters are set as follows: $L_{0} = -30\text{dB}$, $\alpha_{\text{B},i}=\alpha_{\text{B},\text{E}}=4.6$, $\alpha_{\text{B},\text{I}_{l}}=2.2$, $\alpha_{\text{I}_{l},i}=\alpha_{\text{I}_{l},\text{E}}=2.8$, $\kappa=5$, $\sigma^{2}=-105$dBm, $\varrho_{\text{t}}=10^{-5}$ and $\epsilon=0.01$. Furthermore, each point is the average result over $100$ times independent Monte-Carlo trials.

The convergence behaviors of the ring-penalty based SCA and the AO algorithms are evaluated in Fig. \ref{Fig.5}. To illustrate the convergence of AO algorithm, we neglect the inner iteration steps for optimizing the transmit beamforming and reflecting coefficients, and only record the number of the outer alternating iterations. As can be observed, both minimum secrecy rates returned by two algorithms increase monotonically and are guaranteed to converge to the stationary point values within the finite iterations. It is also observed that even adopting the worst-case assumption that E can cancel the co-channel interference of NOMA transmission, the secrecy performance of the AO algorithm outperforms the ring-penalty based SCA algorithm. This is because that: 1) by designing the reasonable beamforming vectors, the multi-antenna BS can fully unleash the spatial degrees of freedom to suppress the co-channel interference at legitimate users, and meanwhile, effectively degrade the received signal power at E, and 2) IRSs lose ability of adjusting eavesdropper's channel in single-antenna case, but have significant inhibitory effects on eavesdropper's channel quality in multi-antenna case.

\begin{figure}[h]
\centering 
\begin{minipage}[b]{0.32\textwidth} 
\centering 
\includegraphics[width=1\textwidth]{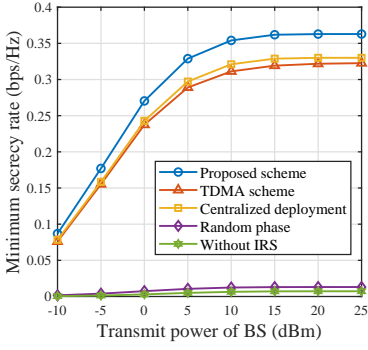} 
\caption{The minimum secrecy rate versus transmit power of BS for single-antenna case with $K=2$, $N=20$, $\text{r}_\text{c}=10$m and $p_{\text{max}}=0.1$.}
\label{Fig.6}
\end{minipage}
\begin{minipage}[b]{0.32\textwidth} 
\centering 
\includegraphics[width=1\textwidth]{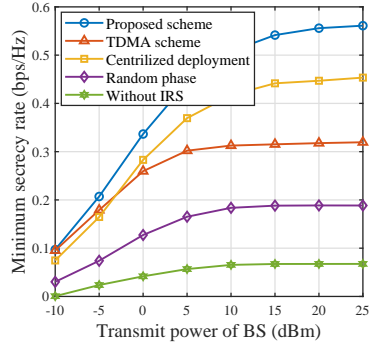}
\caption{The minimum secrecy rate versus transmit power of BS for multi-antenna case with $K=2$, $N=20$, $\text{r}_\text{c}=10$m, $p_{\text{max}}=0.1$ and $M=6$.}
\label{Fig.7}
\end{minipage}
\begin{minipage}[b]{0.32\textwidth} 
\centering 
\includegraphics[width=1\textwidth]{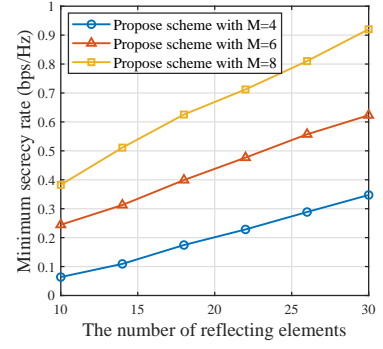} 
\caption{The minimum secrecy rate versus the number of reflecting elements for different number of transmit antennas with $K=2$, $P_{\text{B}}=15$dBm, $\text{r}_\text{c}=10$m and $p_{\text{max}}=0.1$.}
\label{Fig.8}
\end{minipage}
\end{figure}

To demonstrate the performance of the proposed framework, we adopt the following baseline schemes for comparison:
\begin{itemize}
  \item \textbf{Time Division Multiple Access (TDMA)}: In TDMA, the overall transmission phase are equally divided into $K$ orthogonal time slots, where each legitimate user is only scheduled for communication in one time slot. 
  \item \textbf{Centralized deployment}: In centralized deployment, the $N$ passive reflecting elements constitute one IRS, which is located at ($20-\text{r}_\text{c}$, $50$, $0$) m.
  \item \textbf{Random Phase (RP)}: In this case, the phase shifts of reflecting elements are generated randomly in $[0,2\pi)$. The BS optimizes the power allocation or the transmit beamforming according to the combined channels of receiving nodes.
  \item \textbf{Without IRS (WI)}: This scheme neglects the IRS associated links and designs the transmit power/beamforming strategy according to the direct link channels.
\end{itemize}


In Fig. \ref{Fig.6} and Fig. \ref{Fig.7}, we compare the minimum secrecy rate versus the transmit power of BS for different transmission schemes. First, it is observed that the achievable minimum secrecy rate increases gradually with the increasing transmit power, which increases rapidly in the low power regime  while varies slowly in the high power regime. The main reasons are as follows. 1) when the transmit power is low, the receiving signal strength at E is weak, which does not need lots of redundant rate to resist eavesdropping. Accordingly, the transmit power/beamforming and phase shifts mainly focus on enhancing the achievable rate of legitimate users, thus significantly improving the minimum secrecy rate. 2) While when the transmit power becomes large, the receiving signal power at E is strong, which requires a large redundant rate to guarantee the secrecy performance of network. Therefore, even though the achievable rate of legitimate users increases with the increased transmit power, the positive difference between achievable rate and redundant rate, i.e., secrecy rate, is almost unchanged. Second, it can be seen that the proposed NOMA scheme is capable of providing the higher security than the TDMA scheme with the same transmit power, which is due to the fact that the NOMA transmission can serve all the legitimate users simultaneously in the whole transmission phase, which thus improves the secrecy performance of the network. Furthermore, it is also observed that two-IRS distributed deployment scheme achieves better secrecy performance than the centralized deployment scheme, which is because the reflecting elements spread over the distributed IRSs can achieve the higher channel diversity and the joint passive beamforming in a collaborative manner. We refer to this phenomenon as \textit{distance effect}, which is an additionally passive gain brought by the distributed deployment. Finally, it is found that the minimum secrecy rate achieved by the RP and WI schemes are lower than other schemes. This is since that the secrecy performance of the considered network mainly depends on the channel condition gap between legitimate users and E, and the reasonable phase shift design of IRSs can improve the legitimate channels effectively and expand this gap, so as to further improve the network security.

As illustrated in Fig. \ref{Fig.8}, by gradually increasing the number of reflecting elements from $10$ to $30$ with step interval of $4$, the minimum secrecy rate among legitimate users increases monotonically. The reason lies in the following two aspects: 1) the increased number of reflecting elements can establish more reliable cascaded communication links between the BS and receivers and offer the higher array gains; 2) more reflecting elements can provide larger passive beamforming design space, which brings more remarkably passive gains for supporting secure transmission. Besides, we also observe that the minimum secrecy rate increases with the increase of number of transmit antennas $M$. It is because increasing the number of transmit antennas leads to more available spatial degrees of freedom at the BS, which can be fully used by the proposed AO algorithm to secure the legitimate communications.

\begin{figure}[h]
\centering 
\begin{minipage}[b]{0.32\textwidth} 
\centering 
\includegraphics[width=1\textwidth]{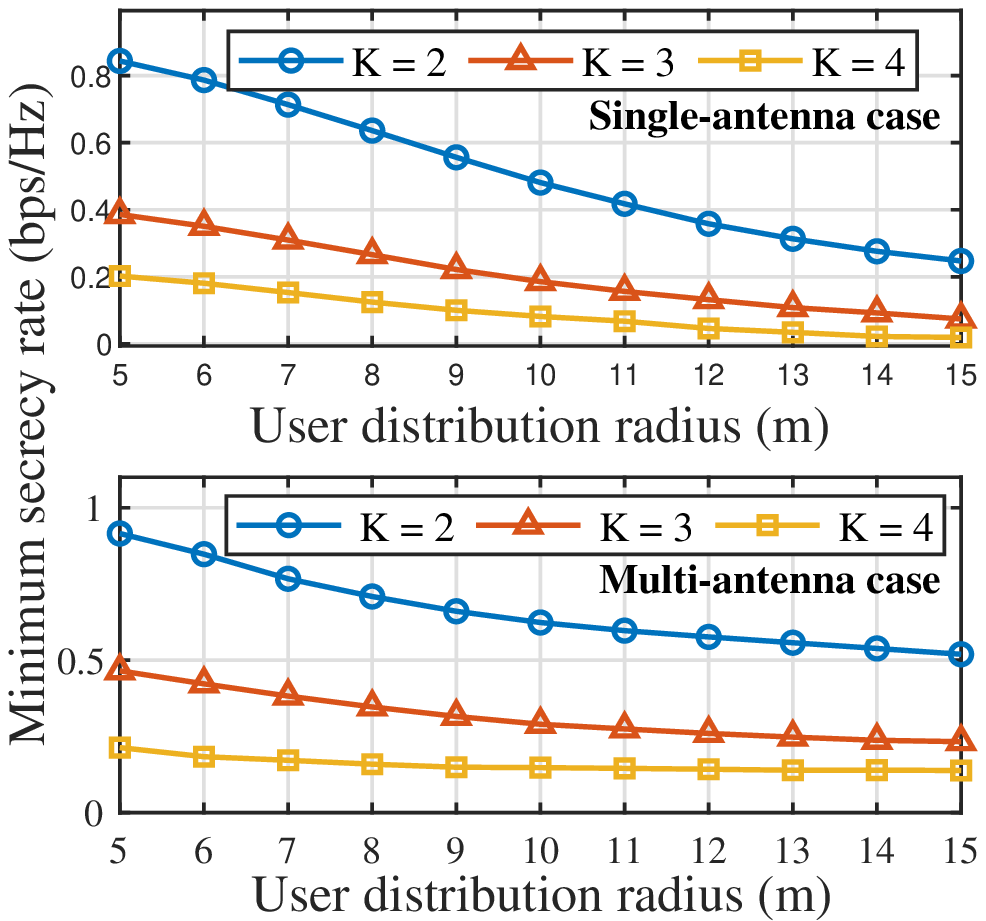}
\caption{The minimum secrecy rate versus the user distribution radius for different number of legitimate users with $N=20$, $P_{\text{B}}=15$dBm, $p_{\text{max}}=0.1$ and $M=6$.}
\label{Fig.9}
\end{minipage}
\begin{minipage}[b]{0.32\textwidth} 
\centering 
\includegraphics[width=1\textwidth]{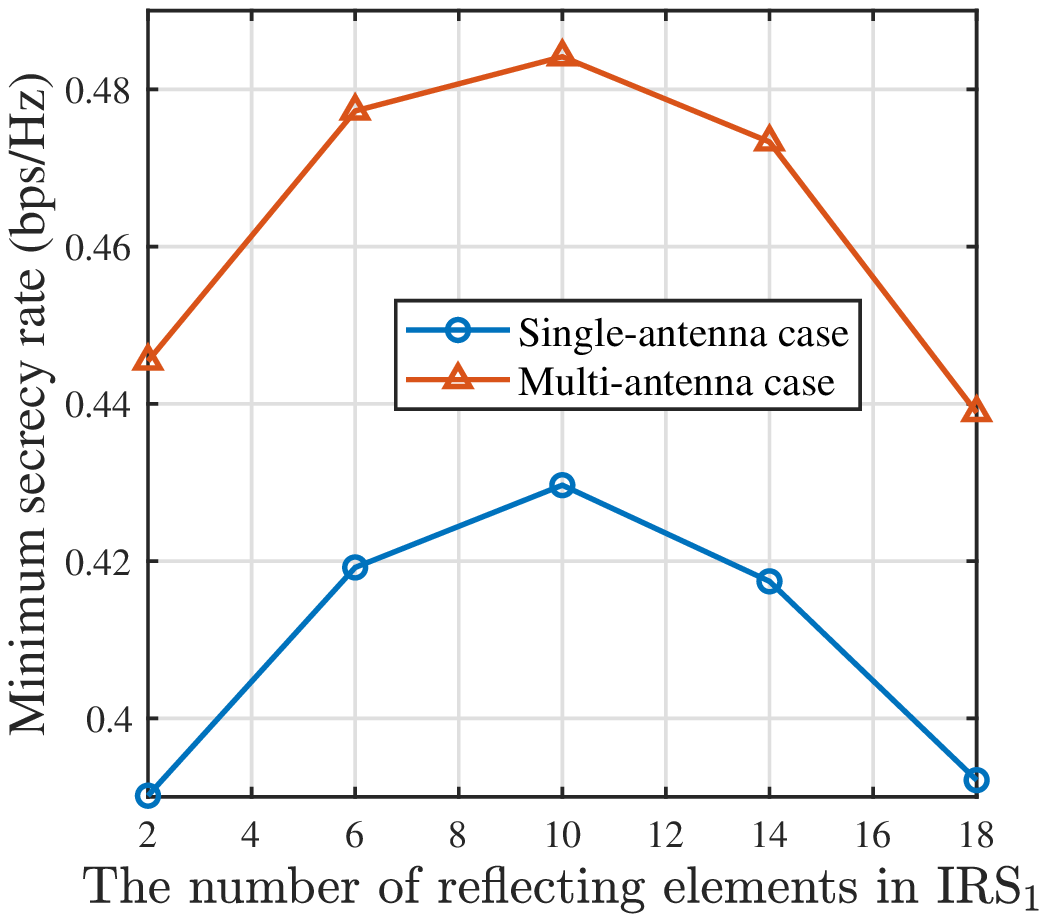} 
\caption{The minimum secrecy rate versus the number of reflecting elements of $\text{IRS}_{1}$ for $K=2$, $N=20$, $\text{r}_{\text{c}}=10$m, $P_{\text{B}}=15$dBm, $p_{\text{max}}=0.1$ and $M=6$.}
\label{Fig.10}
\end{minipage}
\begin{minipage}[b]{0.32\textwidth} 
\centering 
\includegraphics[width=1\textwidth]{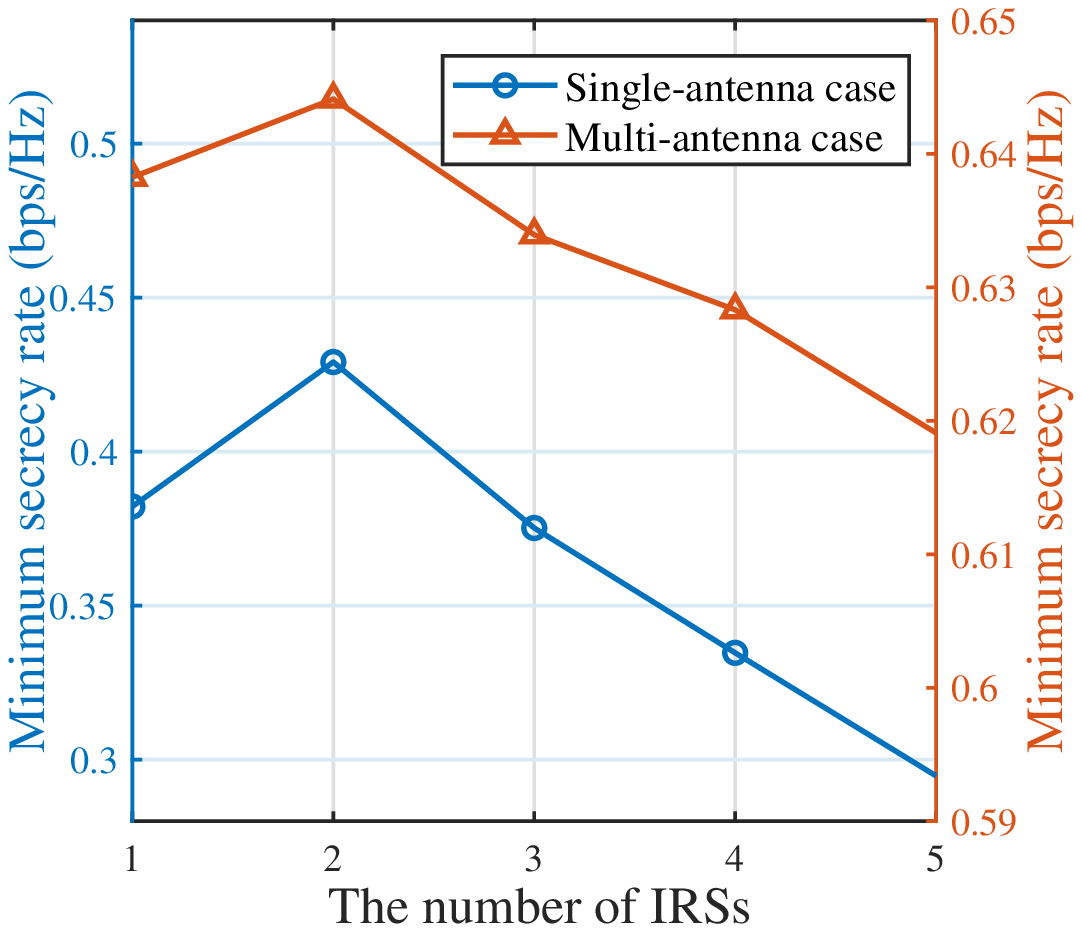}
\caption{The minimum secrecy rate versus the number of IRSs for $K=2$, $N=20$, $\text{r}_{\text{c}}=10$m, $P_{\text{B}}=15$dBm, $p_{\text{max}}=0.1$ and $M=6$.}
\label{Fig.11}
\end{minipage}
\end{figure}


In Fig. \ref{Fig.9}, we investigate the minimum secrecy rate versus the user distribution radius for different number of legitimate users. First, one can observe that the minimum secrecy rate decreases with the increase of the user distribution radius. An intuitive explanation to this phenomenon is that increasing the user distribution radius requires the wider coverage of IRSs, which aggravates the ``double fading'' of the cascade links while weakens the excess passive gains brought by \textit{distance effect}. Then, under the same user distribution radius, we also observe that the secrecy performance of single legitimate user decreases with the increasing user density, which can be expected since that the increase of user density will reduce the power allocated to each user, and thus leads to the minimum secrecy rate reduction at legitimate users.


Fig. \ref{Fig.10} illustrates the minimum secrecy rate versus the number of reflecting elements of $\text{IRS}_{1}$ for two IRSs deployment case. It observed that the minimum secrecy rate increases first and then decreases, which reaches the peak value under the case that $\text{IRS}_{1}$ equips $10$ reflecting elements, i.e., two IRSs share the reflecting elements equally. The reason is that deploying two IRSs with the same number of elements efficiently avoids the situation that one of IRSs is equipped with too few reflecting elements to establish the reliable cascade links while shortens the distance between the reflection elements and the randomly distributed users on average, which fully exerts the passive gains promised by \textit{distance effect}.

In Fig. \ref{Fig.11}, we study the minimum secrecy rate versus the number of IRSs, in which all the IRSs share the reflecting elements equally. Note that for the case of $M=3$, we set $N_{1}=N_{3}=7$ and $N_{2}=6$. As illustrated by Fig. \ref{Fig.11}, it is found that the minimum secrecy rate increases first and then decreases, which achieves the maximum secrecy rate when $L=2$. This result implies that increasing the number of IRSs does not necessarily lead to a higher secrecy performance. In fact, there exists a tradeoff between the number of the IRSs and the number of the reflecting elements equipped at each IRS, which can be further optimized to achieve the optimal performance. Specifically, if we allocate the given number of reflecting elements to more IRSs, each IRS will be equipped with fewer elements, which may not be able to establish the strong links to support wireless communications. On the contrary, if we allocate the reflecting elements to fewer IRSs, the passive gains caused by \textit{distance effect} will be weakened.

\section{Conclusion}\label{conclusion}
This paper investigates the IRS assisted secure NOMA network, where a BS transmits confidential data to the NOMA users with assistance of distributed IRSs against a passive eavesdropper. We aim to maximize the minimum secrecy rate of legitimate users by designing transmit power/beamforming and reflection coefficients jointly, subject to the transmit power constraint at the BS, the phase shifts constraints of IRSs, the SIC decoding constraints and the SOP constraints. For the case with a single-antenna BS, We derive the exact SOP in closed-form expressions and propose a ring-penalty based SCA to optimize the power allocation and phase shifts jointly. Then, we consider the general multi-antenna BS case, and develop a Bernstein-type inequality approximation based AO algorithm to design the transmit beamforming matrix at the BS and optimize the reflection coefficients of IRSs alternately. In particular, we emphasize the difference of the impacts brought by IRSs on the secrecy performance of single-antenna and multi-antenna networks. Numerical results demonstrate the convergence of the proposed algorithms, which achieve better secrecy performance in comparison to other baseline schemes. Also, some practical guidance information of the distributed IRS design is provided.

\section*{Appendix A: Proof of Proposition 1}

To prove proposition 1, we first rewrite the combined channel of E as $h_{\text{E}}=\sum_{n=1}^{N}e^{j\alpha_{n}}h_{\text{B},\text{I}_{n}}h_{\text{I}_{n},\text{E}}^{H}+h_{\text{B},\text{E}}^{H}$. By substituting the results in \cite[Lemma 2]{Z.Ding_IRS_NOMA_2} and the large-scale path losses into the combined channel of E, it follows that $h_{\text{E}}\sim\mathcal{CN}(0,\xi_{\text{E}}^{2}+|L_{\text{B},\text{E}}|^{2})$, with $\xi_{\text{E}}^{2}=\sum_{l=1}^{L}|L_{\text{B},\text{I}_{l}}L_{\text{I}_{l},\text{E}}|^{2}N_{l}$. Note that even though the results in \cite[Lemma 2]{Z.Ding_IRS_NOMA_2} are strictly true when $N_{l}\rightarrow\infty$, the numerical results in \cite{Z.Ding_IRS_NOMA_2} have validated the approximation tightness when $N_{l}$ is small. Thus, it is obvious that $|h_{\text{E}}|^{2}$ follows the exponential distribution, i.e.,

\begin{equation}\label{A1-1}
\tag{A1-1}
f(|h_{\text{E}}|^{2}) = \frac{1}{\xi_{\text{E}}^{2}+|L_{\text{B},\text{E}}|^{2}}e^{-\frac{|h_{\text{E}}|^{2}}
{\xi_{\text{E}}^{2}+|L_{\text{B},\text{E}}|^{2}}}.
\end{equation}
While constraint \eqref{13} can be rewritten as $\mathbb{P}\Bigg(|h_{\text{E}}|^{2}\geq\frac{t_{i}\sigma^{2}}{P_{i}-\sum_{j=i+1}^{K}t_{i}P_{j}}\Bigg) \leq p_{\text{max},i}$, in which the probability operation can be calculated by
\begin{align}\label{A1-2}
\nonumber
\mathbb{P}\Bigg(|h_{\text{E}}|^{2}\geq\frac{t_{i}\sigma^{2}}{P_{i}-\sum_{j=i+1}^{K}t_{i}P_{j}}\Bigg) &= \int\limits_{\frac{t_{i}\sigma^{2}}{P_{i}-\sum_{j=i+1}^{K}t_{i}P_{j}}}^{\infty}
\frac{1}{\xi_{\text{E}}^{2}+|L_{\text{B},\text{E}}|^{2}}e^{-\frac{|h_{\text{E}}|^{2}}{\xi_{\text{E}}^{2}+|L_{\text{B},\text{E}}|^{2}}}
d|h_{\text{E}}|^{2},\\ \tag{A1-2}
&=e^{-\frac{t_{i}\sigma^{2}}{(P_{i}-\sum_{j=i+1}^{K}t_{i}P_{j})(\xi_{\text{E}}^{2}+|L_{\text{B},\text{E}}|^{2})}}.
\end{align}
Therefore, the SOP constraint is given by
\begin{align}\label{A1-3}
\tag{A1-3}
e^{-\frac{t_{i}\sigma^{2}}{(P_{i}-\sum_{j=i+1}^{K}t_{i}P_{j})(\xi_{\text{E}}^{2}+|L_{\text{B},\text{E}}|^{2})}}\leq p_{\text{max},i},
\end{align}
which can be simplified as $t_{i}\geq\frac{\log(\frac{1}{p_{\text{max},i}})(\xi_{\text{E}}^{2}+|L_{\text{B},\text{E}}|^{2})P_{i}}
    {\log(\frac{1}{p_{\text{max},i}})\sum_{j=i+1}^{K}(\xi_{\text{E}}^{2}+|L_{\text{B},\text{E}}|^{2})P_{j}+\sigma^{2}}$. This completes proof.

\section*{Appendix B: Proof of Lemma 1}

Considering the AGM approximation of \eqref{35}, we have inequality $z_{i,i}\leq z_{i,i}^{[\text{max}]}=\frac{\text{Tr}(\mathbf{H}_{i}\mathbf{W}_{i})}{\sum_{j=i+1}^{K}\text{Tr}(\mathbf{H}_{i}\mathbf{W}_{j})+\sigma^{2}}$. While from \eqref{39}, we can derive the lower bound function of secrecy rate, i.e.,
\begin{equation}\label{A2-1}
\tag{A2-1}
f_{s,\text{lower}}=\log_{2}\Big(1+z_{i,i}\Big)-\log_{2}(1+t_{i}),\quad(1\leq i\leq K),
\end{equation}
which is a monotonically increasing function of $z_{i,i}$. Therefore, when the objective reaches optimal value, $f_{s,\text{lower}}$ reaches maximum, which is achieved by $z_{i,i}=z_{i,i}^{[\text{max}]}=\frac{\text{Tr}(\mathbf{H}_{i}\mathbf{W}_{i})}{\sum_{j=i+1}^{K}\text{Tr}(\mathbf{H}_{i}\mathbf{W}_{j})+\sigma^{2}}$. This completes proof.

\section*{Appendix C: Proof of Proposition 2}

In practical transmissions, legitimate users usually possess the limited signal decoding ability, while the potential eavesdropper may have the stronger multi-user detection and interference cancellation capacities. To this end, we adopt the worst-case assumption in PLS \cite{X.Yu_imperfectCSI,ZZ_Robust,H.-M.Wang_NOMA_PLS} that eavesdropper can cancel the co-channel interference in NOMA transmission. Thus, we write the left-hand side of SOP constraint \eqref{34} as
\begin{align}
\nonumber
\text{left-hand side} &=\mathbb{P}\Big((\mathbf{h}_{\text{I},\text{E}}^{H}\bm{\Theta}\mathbf{H}_{\text{B},\text{I}}+\mathbf{h}_{\text{B},\text{E}}^{H})\mathbf{W}_{i}(\mathbf{H}_{\text{B},\text{I}}^{H}\bm{\Theta}^{H}\mathbf{h}_{\text{I},\text{E}}+\mathbf{h}_{\text{B},\text{E}})
>t_{i}\sigma^{2}\Big),\\ \nonumber
&=\mathbb{P}\Big(\mathbf{h}_{\text{I},\text{E}}^{H}\bm{\Theta}\mathbf{H}_{\text{B},\text{I}}\mathbf{W}_{i}\mathbf{H}_{\text{B},\text{I}}^{H}\bm{\Theta}^{H}\mathbf{h}_{\text{I},\text{E}}
+2\Re(\mathbf{h}_{\text{I},\text{E}}^{H}\bm{\Theta}\mathbf{H}_{\text{B},\text{I}}\mathbf{W}_{i}\mathbf{h}_{\text{B},\text{E}})+\mathbf{h}_{\text{B},\text{E}}^{H}\mathbf{W}_{i}\mathbf{h}_{\text{B},\text{E}}
>t_{i}\sigma^{2}\Big),\\ \tag{A3-1}
\label{A3-1}&=\mathbb{P}\Big(\begin{bmatrix}\mathbf{\tilde{h}}_{\text{B},\text{E},\text{s}}^{H} & \mathbf{\tilde{h}}_{\text{I},\text{E},\text{s}}^{H}\end{bmatrix}\bm{\Phi}_{i}
\begin{bmatrix}\mathbf{\tilde{h}}_{\text{B},\text{E}} \\ \mathbf{\tilde{h}}_{\text{I},\text{E}}\end{bmatrix}>t_{i}\sigma^{2}
\Big),
\end{align}
where $\bm{\Phi}_{i}$ is given in \eqref{37}, while $\mathbf{\tilde{h}}_{\text{B},\text{E}}, \mathbf{\tilde{h}}_{\text{I},\text{E}}\sim \mathcal{CN}(0,\mathbf{I})$ denote the small-scale Rayleigh fading channels. Then, by substituting $\mathbf{z}=[\mathbf{\tilde{h}}_{\text{B},\text{E}}, \mathbf{\tilde{h}}_{\text{I},\text{E}}]^{T}$, $\mathbf{A}_{i}=\bm{\Phi}_{i}$ and $x=\log\left(\frac{1}{p_{\text{max},i}}\right)$ into the Bernstein-type inequality \cite[eq. (0.3)]{Bernstein}, we can rewrite the SOP constraint as
\begin{equation}\label{A3-2}
\tag{A3-2}
\mathbb{P}\left(T_{i}\geq \text{Tr}(\bm{\Phi}_{i})+\sqrt{2\log\left(\frac{1}{p_{\text{max},i}}\right)}\|\bm{\Phi}_{i}\|_{F}+\rho_{\text{max}}(\bm{\Phi}_{i})\log\left(\frac{1}{p_{\text{max},i}}\right)\right)\leq p_{\text{max},i},
\end{equation}
where $T_{i}=\mathbf{z}^{H}\mathbf{A}_{i}\mathbf{z}$. Therefore,
$\mathbb{P}\left(T\geq t_{i}\sigma^{2}\right)\leq p_{\text{max},i}$ will hold if the following condition is satisfied
\begin{equation}\label{A3-3}
\tag{A3-3}
t_{i} \geq \frac{1}{\sigma^{2}}\left( \text{Tr}(\bm{\Phi})+\sqrt{2\log\left(\frac{1}{p_{\text{max},i}}\right)}\|\bm{\Phi}\|_{F}+\rho_{\text{max}}(\bm{\Phi})\log\left(\frac{1}{p_{\text{max},i}}\right)\right).
\end{equation}
 To tackle the non-convex operation $\rho_{\text{max}}(\bm{\Phi}_{i})$, we introduce the auxiliary variable $\Phi_{i}$ to replace the maximal eigenvalue of $\bm{\Phi}$, which equivalently transforms \eqref{A3-3} into
\begin{subequations}
    \begin{gather}\label{A3-4a}
    \tag{A3-4a}t_{i}\geq \frac{1}{\sigma^{2}}\left( \text{Tr}(\bm{\Phi}_{i})+\sqrt{2\log\left(\frac{1}{p_{\text{max},i}}\right)}\|\bm{\Phi}_{i}\|_{\text{F}}+\log\left(\frac{1}{p_{\text{max},i}}\right)\phi_{i}\right),\quad (1\leq i\leq K), \\ \tag{A3-4b}
    \label{A3-4b} \phi_{i}\mathbf{I}-\bm{\Phi}_{i}\succeq \mathbf{0},\quad (1\leq i\leq K).
    \end{gather}
\end{subequations}
This completes proof.


\end{document}